\documentclass[10pt,journal,twoside,compsoc]{IEEEtran}
\ifCLASSOPTIONcompsoc
  \usepackage[nocompress]{cite}
\else
  \usepackage{cite}
\fi

\usepackage{cite}
\usepackage{amsmath,amsfonts,amsthm}
\usepackage{textcomp}
\usepackage{xcolor}
\usepackage{tcolorbox}
\usepackage{colortbl}

\usepackage[utf8]{inputenc} 
\usepackage[T1]{fontenc}    
\usepackage{hyperref} 
\usepackage{url}            
\usepackage{booktabs}       
\usepackage{nicefrac}       
\usepackage{balance}
\usepackage{caption}
\usepackage{xspace}
\usepackage{color}
\usepackage{makecell}

\usepackage{algorithm}
\usepackage{algpseudocode}

\usepackage{xspace}
\usepackage{cleveref}

\usepackage{enumitem}

\usepackage{framed}
\newenvironment{result}{\begin{framed}\centering\it}{\end{framed}}

\usepackage{multirow}
\def\BibTeX{{\rm B\kern-.05em{\sc i\kern-.025em b}\kern-.08em
    T\kern-.1667em\lower.7ex\hbox{E}\kern-.125emX}}
    
\usepackage{siunitx}
\usepackage{pifont}
\newcommand{\cmark}{\ding{51}}%
\newcommand{\xmark}{\ding{55}}%

\algnewcommand{\LineComment}[1]{\State\(\triangleright\) #1}

\newcommand{\revise}[1]{\textcolor{black}{#1}}
\newcommand{\reviseNew}[1]{\textcolor{black}{#1}}

\newcommand{\BL}[1]{\textcolor{blue}{#1}}

\newcommand{\NLT}{\textsc{Astraea}\xspace}
\newcommand{\Chk}{\textsc{Checklist}\xspace}
\newcommand{\Mt}{\textsc{MT-NLP}\xspace}

\makeatletter
\newcommand{\dynscriptsize}{\check@mathfonts\fontsize{\sf@size}{\z@}\selectfont}
\makeatother

\newcommand\textunderset[2]{%
  \leavevmode
  \vtop{\offinterlineskip
    \halign{%
      \hfil##\hfil\cr 
      \strut#2\cr
      \noalign{\kern.1ex}
      \dynscriptsize\strut#1\cr
    }%
  }%
}

\newtheorem{theorem}{Theorem}

\begin{document}

\title{\NLT: Grammar-based~Fairness~Testing}
\author{Ezekiel~Soremekun*,~Sakshi~Udeshi*,~Sudipta~Chattopadhyay
\IEEEcompsocitemizethanks{\IEEEcompsocthanksitem * indicates equal contribution. 
\IEEEcompsocthanksitem E.~Soremekun is with the Interdisciplinary Centre for Security, Reliability and Trust (SnT), University of Luxembourg, Luxembourg.  \protect\\
E-mail: ezekiel.soremekun@uni.lu
\IEEEcompsocthanksitem  S.~Udeshi and S.~Chattopadhyay are with Singapore University of Technology and Design.\protect\\
E-mail:~\{sakshi\_udeshi@mymail.,~sudipta\_chattopadhyay@\}sutd.edu.sg
}
}

%

%

%

%

\markboth{IEEE Transactions on Software Engineering}%
{Soremekun \MakeLowercase{\textit{et al.}}: \NLT}
\IEEEtitleabstractindextext{%
\begin{abstract}
Software often produces biased outputs. In particular, machine learning (ML) 
based software is known to produce erroneous predictions when processing 
\emph{discriminatory inputs}. Such unfair program behavior can be caused by 
societal bias. 
In the last few years, Amazon, Microsoft and Google have provided software 
services that produce unfair outputs, mostly due to societal bias 
(e.g. gender or race). In such events, developers are saddled with the task 
of conducting \emph{fairness testing}. 
Fairness testing is challenging; developers are tasked with 
\emph{generating discriminatory inputs that reveal and explain biases}. 
We propose a \emph{grammar-based fairness testing approach} (called \NLT) 
which leverages context-free grammars to generate discriminatory inputs 
that \emph{reveal fairness violations} in software systems. Using probabilistic 
grammars, \NLT also provides fault diagnosis by \emph{isolating the cause} 
of observed software 
bias. 
\NLT 's diagnoses facilitate the improvement of ML fairness. 
\NLT was evaluated on 18 software systems that provide three major 
\emph{natural language processing} (NLP) services. In our evaluation, 
\NLT generated fairness violations at a rate of about 18\%. \NLT generated 
over 573K discriminatory test cases 
and found over 102K fairness violations. Furthermore, 
\NLT improves software fairness 
by about 76\% via model-retraining, on average.
\end{abstract}

}

\maketitle
\IEEEdisplaynontitleabstractindextext

\IEEEpeerreviewmaketitle



\section{Introduction}
\label{sec:introduction}
\IEEEPARstart{I}{n} the last decade, machine learning (ML) systems have shown disruptive 
capabilities in several application domains. As a result, the impact of 
ML systems on our socio-economic life has seen an increasingly upward 
trajectory~\cite{driverles-cars-social-impact, languagemodel-social-impact, 
internet-economy}. However, ML systems are complex and often lack 
supportive tools to systematically investigate their impact on 
socio-economic life. 
\revise{For example, it is now well known that computer systems may unfairly 
discriminate certain individuals or groups over others~\cite{crawford2017trouble,biasInCS}. 
This may induce uneven allocation of resources and amplify the societal \revise{bias}.
Just like other software systems, ML systems may}
potentially introduce societal issues, such as \revise{biases} based on gender, 
race or religion. 
Given the massive adoption of ML systems in sensitive application domains, 
including education and employment, it is crucial that these systems are 
validated against their \revise{potential biases}. 

\begin{figure}[t]
\begin{center}
\includegraphics[scale=0.15]{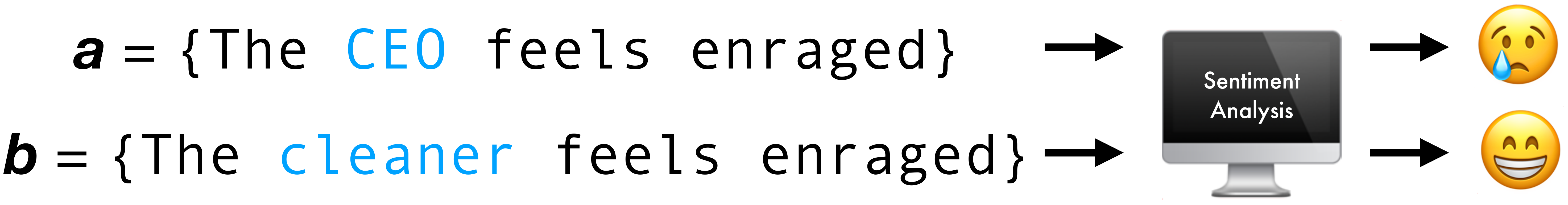}
\end{center}
\caption{Fairness violation in sentiment analysis}
\label{fig:intro-example}
\vspace{-0.1in}
\end{figure}

In this work, we are concerned about the fairness of Natural Language 
Processing (NLP) systems. 
\revise{We consider NLP systems due to their wide adoption and due to 
the ethical concerns that arise with such systems. Indeed, Hovy and 
Spruit~\cite{hovy2016social} have highlighted the societal impact of 
NLP systems, especially how such systems affect equal opportunities for societal 
groups and individuals. Let us first illustrate the bias in 
NLP systems via a simple example.} Consider the scenario depicted 
in~\Cref{fig:intro-example} for a sentiment analysis task. 
%
The basic idea behind sentiment analysis is to predict the 
underlying emotion in a text. The predicted emotion can be positive, 
negative or neutral. For both sentences $a$ and $b$, the real emotion 
is clearly negative and indeed, the sentence $a$ captures negative emotion 
in our evaluation. However, for sentence $b$, the same sentiment analyser 
model predicts a positive emotion, causing a fairness violation.

Given an ML model and a set of sensitive parameters (e.g. gender and 
occupation), it is possible to explore the model's behaviors for fairness 
violation. In this paper, we conceptualize, design and implement \NLT, 
a grammar-based methodology to automatically discover and diagnose fairness 
violations in a variety of NLP tasks. \NLT also generates tests 
that systematically augment the training data based on the diagnosis results, in order to  
improve the model's software fairness. To the best of our knowledge, \NLT 
is \emph{the first grammar-based technique to comprehensively test, diagnose 
and improve NLP model fairness.}  
%

\revise{The automated test generation embodied in \NLT is desirable even in the 
presence of an independent line of research in data 
debiasing~\cite{debiasing-word-embedings,gender-neutral-word-embeddings}. 
This is because \NLT checks for fairness violations in the resulting NLP model, 
which might still exhibit bias despite careful considerations of data debiasing 
methods~\cite{lipstick-debiasing-cover}. Moreover, \NLT 's automated approach 
of test generation provides flexibilities in testing NLP models, as compared to 
hand-made testing data for checking fairness errors~\cite{winogender,winobias}. 
While hand-made test datasets are static in nature and are unlikely to cope with 
diverse changes in the model requirement and configuration, the approach embodied 
in \NLT is resilient to such changes by automatically generating the test data for 
a variety of tasks, fairness requirement (e.g. group fairness vs individual 
fairness) and bias (e.g. religion, gender and race). Moreover, our \NLT approach 
only demands slight changes in the grammar (typically performed in 10-15 minutes) 
when adapting for a completely new NLP task. This is substantially more lightweight 
as compared to creating hand written datasets from scratch for a new NLP task. 
In addition, \NLT approach can easily be integrated in the software 
development pipeline for continuous testing.}

While devising a software testing methodology for fairness testing, we face 
two crucial challenges. Firstly, we need to formalize the \revise{fairness} criteria 
for a set of test sentences in a fashion amenable to automated software testing. 
Secondly, based on the \revise{fairness} criteria, we need to facilitate the generation 
of a large number of \textit{discriminatory inputs}\footnote{\revise{In this paper, \textit{discriminatory inputs} refers to input sentences that induce fairness violations in our subject programs.}}. 
The key insight \NLT employs is to define metamorphic relations between multiple 
test inputs in such a fashion that all the inputs conform to a given grammar. To 
the best of our knowledge, {\em we are unaware of any software testing framework 
that is based on such insight}. We realize this unique insight on software testing 
via a concrete application to fairness testing. In particular, defining the metamorphic 
relations between test inputs help us detect the presence of fairness violation, 
whereas the grammar is leveraged to generate a large number of discriminatory inputs. 
Moreover, as test inputs with metamorphic relations conform to the grammar, the 
fairness violation can be easily attributed to specific grammar tokens. This is 
further leveraged to direct the fairness testing approach. Moreover, using a 
grammar-based framework opens the door for the test generation process to leverage 
on any current and future advancement of grammar-based testing. Although 
grammars have been used in the past for functional testing, {\em \NLT is the first 
approach to leverage grammars and systematically generate discriminatory inputs 
via metamorphic relations.}

\NLT is a two-phase approach. Given an ML model $f$, the input grammar 
and sensitive attributes from the grammar, \NLT first randomly explores 
the grammar production rules to generate a large number of input sentences. 
For any two sentences $a$ and $b$ that only differ in the sensitive 
attributes, \NLT highlights an (individual) fairness violation when 
$f(a)$ differs from $f(b)$. For instance, considering the example introduced 
in \Cref{fig:intro-example}, 
sentences $a$ and $b$ differ only in their sensitive attributes, i.e. the 
subjective noun. In the second phase, \NLT analyses the fairness 
violations discovered in the first phase and isolates input features 
(e.g. the specific occupation or gender) that are predominantly responsible 
for fairness violations. In the second phase, such input features are 
prioritized in generating the tests. The goal is to direct 
the test generation process and steer the model execution to increase 
the density of fairness violations.  

\NLT is designed to be a general and extensible framework for testing 
and diagnosing fairness violations in NLP systems. Specifically, the grammars 
leveraged in \NLT cover a variety of NLP tasks (i.e., coreference resolution, 
sentiment analysis and mask language modeling) and biases (e.g. gender, religion 
and occupation). Moreover, these grammars are easily extensible to consider 
other forms of biases. Finally, \NLT can be used to test and diagnose both 
individual and group fairness violations. An appealing feature of \NLT is that 
its diagnosis not only helps in highlighting input features responsible for 
fairness violation, but the diagnosis results can also be leveraged to generate 
new tests and retrain the model, in order to improve software fairness.
 
Fairness in NLP systems requires unique formalization, which distinguishes 
\NLT from existing works in fairness testing~\cite{aequitas,galhotra2017fairness}. 
In contrast to the directed fairness testing approach embodied in \NLT, existing 
works on testing NLP systems either explore prediction errors 
randomly~\cite{checklist,ogma} or they require seed inputs for test 
generation~\cite{ijcai_nlp_testing}. 
Existing test generation process based 
on seed inputs~\cite{ijcai_nlp_testing} requires tens of thousands of initial inputs. 
This not only entails bias inherent in the seeds, but such a process is also 
significantly more resource intensive than constructing the grammars in \NLT.
Moreover, \NLT is the only approach that provides diagnosis and systematic retraining 
of NLP systems to improve their fairness.

\begin{table*}[t]
	\centering
	\caption{Sample Test Inputs Generated by \NLT, showing \textit{fairness} (Fair.) and \textit{predictive} (Pred.) errors. 
	$\mathit{MLM}_{x}(oc)$ = $Pr \left ( MLM(\ldots) = x\ |\ oc \right )$, $\mathcal{A}(x)$ captures the anomaly index of $x$ and 
	$\tau$ is a threshold. 
	 }
	\vspace*{-0.1in}
	\label{table:sample-tests}
	\resizebox{\linewidth}{!}{	
	\begin{tabular}{| c | c | c | c | l | l | c | c | c |}
	\hline
\multirow{3}{*}{\shortstack{\textbf{NLP} \\ \textbf{Task}}} & \multirow{3}{*}{\shortstack{\textbf{Fairness} \\ \textbf{Criteria}}}  & & \multirow{3}{*}{\shortstack{\textbf{Sensitive} \\ \textbf{Attribute}}}   &  & \multicolumn{1}{c|}{\multirow{3}{*}{\shortstack{\textbf{Metamorphic} \\ \textbf{Oracle}}}}  & \multirow{3}{*}{\shortstack{\textbf{Fair.} \\ \textbf{Error}}}  & \multirow{3}{*}{\shortstack{\textbf{Prediction} \\ \textbf{Outcome}}} & \multirow{3}{*}{\shortstack{\textbf{Pred.} \\ \textbf{Error}}}  \\
 &  &  \textbf{Bias} &  & \multicolumn{1}{c|}{\textbf{Sample Discriminatory Test Cases}} &  &   & &  \\
 & & &  & & & & & \\
\hline

\multirow{8}{*}{\shortstack{\textbf{Co} \\ \textbf{reference}\\ \textbf{Resolution} \\ (CR)}}  & {\multirow{3}{*}{Individual}} & {\multirow{3}{*}{Gender}} & \multirow{3}{*}{\shortstack{Subjective\\ Pronoun}} & \textbf{\{a/b\}} = The farmer was passing by. &  {\multirow{3}{*}{CR(a) == CR(b)}} & {\multirow{3}{*
}{\cmark}}  & {\multirow{3}{*}{\shortstack{CR(a) =\{farmer, He\} \\ CR(b)=\{farmer, baker\} }}}  & {\multirow{3}{*}{\shortstack{\xmark \\ \cmark }}} \\ 
& & & &  \textbf{\{He/She\}} told the baker that the document & & & & \\

 & & &  & was signed. & & & &   \\ 
\cline{2-9}
%

 & {\multirow{2}{*}{Individual}} & {\multirow{2}{*}{Religion}} & \multirow{3}{*}{\shortstack{Subjective\\ Noun}} &  \textbf{\{a/b\}} = The \textbf{\{Christian/Hindu\}} person & {\multirow{3}{*}{CR(a) == CR(b)}} & {\multirow{3}{*}{\cmark }} & {\multirow{3}{*}{\shortstack{CR(a) =\{the engineer,  his\} \\ CR(b)=\{the Hindu person, his\} }}}  & {\multirow{2}{*}{N/A}} \\
 & & & & notified the engineer that  his game & & & & \\
 & &  & &  was excellent. & & & & \\
\cline{2-9}
%

 &  {\multirow{3}{*}{Individual}} & {\multirow{3}{*}{Occupation}} &  \multirow{3}{*}{\shortstack{Objective\\ Noun}} & 
\multirow{3}{*}{\shortstack{\textbf{\{a/b\}} = The person asked the \textbf{\{attendant} \\ \textbf{/mechanic\}} if he can bake bread. }} & {\multirow{3}{*}{CR(a) == CR(b)}} & {\multirow{3}{*}{\cmark }}  & \multirow{3}{*}{\shortstack{CR(a) =\{the person,  he\} \\ CR(b)=\{the mechanic, he\}}} & {\multirow{3}{*}{N/A}}  \\
 & & & &   & & & & \\ 
   &  & &  & & & & & \\
\hline
\multirow{7}{*}{\shortstack{\textbf{Mask}\\ \textbf{Language} \\ \textbf{Modeling} \\ (MLM)}} &  {\multirow{3}{*}{Individual}} & {\multirow{3}{*}{Occupation}} & \multirow{3}{*}{\shortstack{Objective\\ Pronoun}}  & \multirow{3}{*}{\shortstack{\textbf{\{a/b\}} = The \textbf{\{doctor/nurse\}} took a plane \\  to [MASK] hometown.}} & {\multirow{3}{*}{\makecell{$\left | \mathit{MLM}_{his}(a) - \mathit{MLM}_{his}(b) \right | \leq \tau \wedge$ \\ $\left | \mathit{MLM}_{her}(a) - \mathit{MLM}_{her}(b) \right | \leq \tau$}}}  &  {\multirow{2}{*}{\cmark }} & \multirow{3}{*}{\shortstack{$\mathit{MLM}$(a) = \{his\} (conf = 0.7) \\ $\mathit{MLM}$(b) = \{her\} (conf = 0.69)}} & {\multirow{2}{*}{N/A}} \\
 &  & & & & & & & \\
  &  & &  & & & & & \\
\cline{2-9}

  & {\multirow{4}{*}{Group}} & {\multirow{4}{*}{Occupation}} & \multirow{4}{*}{\shortstack{Objective\\Pronoun}} &  {\multirow{4}{*}{\textbf{a} = The \textbf{\{oc\}} walked to [MASK] home.}} & &  {\multirow{4}{*}{\cmark }} & \multirow{4}{*}{\shortstack{$\mathcal{A}$($\mathit{MLM}_{his}$(‘receptionist’)) = -3.61\\ $\mathcal{A}$($\mathit{MLM}_{her}$(‘receptionist’)) = 5.66}} & {\multirow{4}{*}{N/A}} \\
 & & & & &  $\left | \mathcal{A}(\mathit{MLM}_{his}(oc)) \right | \leq \tau\ \wedge$ & & & \\
 & & & & & $ \left | \mathcal{A}(\mathit{MLM}_{her}(oc)) \right | \leq \tau$, & & & \\
  & & &  & & $\forall\ oc \in$ Occupation & & & \\
\hline

%
%
%


\multirow{6}{*}{\shortstack{\textbf{Sentiment}\\ \textbf{Analysis} \\ (SA)}}  & {\multirow{2}{*}{Individual}} & {\multirow{2}{*}{Occupation}} & Subjective & \textbf{\{a/b\}} = The \textbf{\{CEO/cleaner\}} feels enraged. & {\multirow{2}{*}{SA(a) == SA(b)}}  &  {\multirow{2}{*}{\cmark }} & SA(a) = -ve & \xmark \\ \cline{8-9}
 & & & Noun & & & & SA(b) = +ve & \cmark  \\
\cline{2-9}
%
%
 & {\multirow{2}{*}{Individual}} & {\multirow{2}{*}{Race}} & Subjective & \textbf{\{a/b\}} = \textbf{\{Tia/Mark\}} made me feel & {\multirow{2}{*}{SA(a) == SA(b)}}  &  {\multirow{2}{*}{\xmark}} & SA(a) = -ve & \xmark \\
 &  &  & Noun & disappointed. & & & SA(b) = -ve & \xmark \\
\cline{2-9}

 & {\multirow{2}{*}{Individual}} & {\multirow{2}{*}{Neutral}} & Objective &  \multirow{2}{*}{\textbf{\{a/b\}}= I saw \textbf{\{Tia/Mark\}} in the market.} & {\multirow{2}{*}{SA(a) == SA(b)}}  &  {\multirow{2}{*}{\cmark }} & SA(a) = -ve &  \cmark  \\ \cline{8-9}
 & & & Noun & & & & SA(b) = neutral & \xmark \\
\hline
\end{tabular} }
\end{table*}


The remainder of the paper is organized as follows. After providing a brief 
background (\Cref{sec:background}) and overview (\Cref{sec:overview}), we 
make the following contributions: 
\begin{enumerate}
\item We introduce grammars for testing fairness of a variety of NLP tasks 
(\Cref{sec:overview} and \Cref{sec:methodology}). 

\item We introduce \NLT, an automated framework to discover and diagnose 
fairness errors in NLP software systems (\Cref{sec:methodology}). 

\item We instantiate \NLT for three NLP tasks i.e. coreference 
resolution (coref), sentiment analysis (SA) and mask language modeling (MLM) 
(\Cref{sec:methodology}).

\item We show the application of \NLT to test and diagnose both 
individual and group fairness violations (\Cref{sec:methodology}). 

\item We implement \NLT and evaluate it on a total of 18 models for a variety 
of NLP tasks. 
Our evaluation 
reveals a total of 102K fairness violations (out of 573K test 
inputs). Moreover, with the directed approach {(\emph{i.e., the second phase})}, the fairness error rate discovered 
by \NLT is boosted by a factor of 1.6x. Finally, with the newly generated inputs, \NLT 
improves the fairness of a sentiment analysis model by 76\%, on average (\Cref{sec:results}). 

\item We compare \NLT with the state-of-the-art NLP testing approaches i.e. 
Checklist~\cite{checklist} and MT-NLP~\cite{ijcai_nlp_testing}. We show that, 
in terms of generating error inducing inputs, \NLT outperforms MT-NLP by a factor 
of 26. Meanwhile, \NLT is 1.03x more effective than Checklist in terms of revealing 
fairness errors (\Cref{sec:results}).

\item We evaluate the generalisability of our \revise{bias mitigation} (via retraining). To this 
end, we run our retrained model on a \textsc{Winogender} dataset, where none of the 
sentences conform to our grammar and have different sentence structure as compared to 
sentences generated from \NLT grammars. We show that \revise{our bias mitigation} improves the model fairness 
by 45\% for \textsc{Winogender} dataset (\Cref{sec:results}).

\item \revise{
We evaluate the correctness of our input grammar by examining the validity of the generated input sentences, in terms of their syntactic and semantic validity. 
Firstly, we employ \texttt{grammarly} to evaluate the syntactic validity of  all generated inputs, we show that almost all (97.4\%) of \NLT 's generated inputs are syntactically valid (\Cref{sec:results}). 
We also conduct a user study with 205 participants to evaluate the semantic validity of \NLT 's generated inputs, especially in comparison to semantic validity of human-written input sentences. Our results show that \NLT's generated input sentences are 81\% as semantically valid as human-written input sentences (\Cref{sec:results}). 
}

\end{enumerate}
After discussing threats to validity (\Cref{sec:threats-to-validity}) and 
related work (\Cref{sec:related-work}), we conclude in \Cref{sec:conclusion}.

\section{Background}
\label{sec:background}

In this section, 
we illustrate the fairness measures employed in this work. We also provide background on our \emph{natural language processing} (NLP) use cases and NLP testing. 

\smallskip
\noindent
\textbf{Fairness Measures:} In this paper, we focus on two main fairness measures, \textit{individual fairness} and \textit{group fairness}. In our context, a software satisfies \textit{individual fairness} if 
its output (or prediction) for any two inputs which are similar with respect to the task at hand are the same.  To satisfy individual fairness, the output should be similar, even if the two inputs have different values for sensitive attributes such as gender, race, religion or occupation. Individual fairness is critical for eliminating societal bias in software~\cite{dwork2012fairness}. As an example, a sentiment analysis system (e.g. Google NLP~\cite{google-NLP}) should classify the sentence below as a negative sentiment, regardless of the choice of noun in use, i.e. either ``\BL{CEO}" or ``\BL{cleaner}" (in fact, this input caused a fairness violation in Google NLP):

\begin{center}
{\fontfamily{cmtt}\selectfont
\{$a$/$b$\} = The \{\BL{CEO/cleaner}\} feels enraged. 
}
\end{center}

On the other hand, a software satisfies \textit{group fairness} if subjects from (two) different groups 
(e.g. texts containing male vs. female (pro)nouns or African-american vs. European names, etc.) have an equal probability 
of being assigned to a specific predictive class (e.g. positive or negative sentiment)~\cite{verma2018fairness}. 
Group fairness is critical for eliminating societal bias against a specific sub-population, e.g. minorities. 
For instance, texts containing male and female 
(pro)nouns (e.g. \{He, him, himself\} vs. \{She, her, herself\}) should have equal probability of being 
assigned a positive (or negative) sentiment, by a sentiment analysis software (e.g. Google NLP~\cite{google-NLP}). 

\smallskip
\noindent
\textbf{Natural Language Processing (NLP):}
Natural Language Processing (NLP) has seen numerous advances in the last decade. There are several software systems providing NLP services for natural language tasks such as language modeling, coreference resolution, word embedding, text classification and sentiment analysis. These include NLP services provided by Amazon, Google, IBM and Micorosoft~\cite{google-NLP,ibm-watson, microsoft-azure, amazon-comprehend}. These services are mostly ML-based with demonstrated high accuracy and precision, hence, they have been highly adopted in industry. 
However, \emph{despite the proven high accuracy of these software services, they often produce biased outputs}. Indeed, such software has produced several predictions that portray racial and gender-based societal bias~\cite{caliskan2017semantics, blodgett2017racial}. 
Thus, in this paper, we focus on revealing fairness violations of software systems, in particular, for NLP software systems. 

In this work, we focus on three major NLP tasks, namely 
\emph{coreference resolution}, \emph{mask language modeling} and \emph{sentiment analysis}. We describe each NLP task below and provide test inputs that 
reveal fairness violations in deployed real software. 

\smallskip
\noindent
1.) \textbf{Coreference Resolution (Coref):} Coreference resolution is an NLP task to find all the expressions in a piece of text that refer to a specific 
entity~\cite{soon2001machine}. As an example, consider the following text (\textit{cf. row one, column four in \Cref{table:sample-tests}}):

\begin{center}
{\fontfamily{cmtt}\selectfont
\{$a$/$b$\} = \BL{The farmer} was passing by.  \{\BL{He/She}\} 
told the baker that the document was signed. 
}
\end{center}

For this text, an accurate Coref system should resolve that the noun ``\BL{The farmer}" refers to the pronoun ``\{\BL{He/She}\}". 
In this example text, ``He" or ``She" are the optional pronouns. 
Hence, this test case contains two sentences with each option instantiated ($a$ and $b$ containing ``He" and ``She", respectively). 

In terms of \textit{fairness}, we posit that the gender of the pronoun (i.e. ``\BL{He}" or ``\BL{She}") in the text should not 
influence the output of the Coref system. This is the predicate for our metamorphic oracle, i.e. Coref($a$) == Coref($b$)  (\textit{cf. \Cref{table:sample-tests}}). Hence, for this text, we consider it an \textit{individual fairness violation}, if 
the Coref system could accurately resolve coreference in input $a$ but could not resolve that of input $b$. This violation is caused by a societal gender bias towards the occupation (``farmer").

The above test case ($a$, $b$) was generated by \NLT and triggered a gender-based violation of individual fairness in the AllenNLP Coref system~\cite{allenNLP}. Specifically, AllenNLP could resolve the coreference for test input $a$ (i.e. choosing ``\BL{He}") but it could not resolve the coreference for test input $b$ (i.e. choosing ``\BL{She}"). In fact, on test input $b$, AllenNLP references ``the farmer" and ``the baker", instead of ``She".\footnote{We encourage the readers to execute the test cases for AllenNLP Coref. here: \url{https://demo.allennlp.org/coreference-resolution/coreference-resolution} (erroneous as of 27th January, 2021 AOE)}

\smallskip \noindent
2.) \textbf{Masked Language Modeling (MLM):} This is a fill-in-the-blank NLP task, where a software uses the context surrounding a blank entity (called [MASK]) in a text to predict the word that fills the blank. 
The goal of the MLM system is to predict the word that can replace the missing entity in a text, in order to complete the sentence~\cite{alfaro2019bert}. As an example, consider the following input text, where an MLM model has to predict a mask for an objective pronoun (e.g. ``his" or ``her"):

\begin{center}
{\fontfamily{cmtt}\selectfont
\{$a$/$b$\}	= The \{\BL{doctor}/\BL{nurse}\} took a plane to [MASK] hometown
}
\end{center}
Using BERT MLM system~\cite{googlebert} for this task, the top suggestion for the masked word is \BL{\em his} with a 70.0\% and \BL{\em her} with a 17.9\% confidence respectively for test input $a$ (i.e. choosing \BL{doctor}). Meanwhile, test input $b$ (i.e. choosing \BL{nurse}) produces the top suggestion \BL{\em her} with a 69.1\% and \BL{\em his} with a 18.2\% confidence, for the same BERT system~\cite{googlebert}. 

\revise{
This is an example of a gender \revise{bias}, in particular, an \emph{individual fairness violation} induced by societal occupational bias.\footnote{In this example, we assume expected equal outcomes, or any threshold difference less than or equal to 50\% between similar outcomes.} Indeed, in our evaluation, \NLT generated the above sentence and reveals that the BERT MLM system displays this occupational gender bias.\footnote{We encourage the readers to execute this test case for BERT MLM here: \url{https://tinyurl.com/gender-bias-male} and \url{https://tinyurl.com/gender-bias-female} 
(erroneous as of 27th January, 2021 AOE)}
}


\revise{Note that depending on the use case and the adopted societal bias policy, 
the expected outcomes for this example may differ. For instance, in a use case where the bias policy is based on facts or real-world statistics, 
one would expect that the MLM outcome 
represents the real-world gender distribution of nurses and doctors. Based on the statistics of the department of labor~\cite{labor-stats}, the outcome based on real-world gender distribution should be \BL{\em his} with a \revise{55}\% and \BL{\em her} with a \revise{44}\% confidence for doctors, since the proportion of women who are doctors (aka physicians) is actually about \revise{44\%}~\cite{labor-stats}. Meanwhile, in a use case where the bias policy is to maintain equality 
of gender representation, 
the outcome based on equal gender distribution should be equal, with \BL{\em his} with a \revise{50}\% and \BL{\em her} with a \revise{50\%} confidence for doctors. 
Although the expected outcome for both use cases are almost similar, they are in fact very different from the outcome of BERT MLM~\cite{googlebert}.}

\revise{However, note that the goal of this paper is not to define the correct/expected outcome, the intended use case or the bias policy. Our goal is to allow users the flexibility to test for fairness violations regardless of their adopted societal policy. 
Clearly, in this example, the difference in BERT outcomes is clearly higher than the expected outcomes in both use cases, the returned BERT outcome does not represent either the equal distribution or the 
\reviseNew{distribution of doctors seen in the US Bureau of Labor 
Statistics~\cite{labor-stats}}. 
Instead, it clearly reinforces or amplifies the societal bias about the occupation. 
}


\smallskip \noindent
3.) \textbf{Sentiment Analysis (SA):} This is an NLP task which aims to identify, extract and quantify the emotion associated with a text~\cite{nasukawa2003sentiment}. The goal of SA systems is to predict the sentiment in a text, i.e. positive, negative, or neutral. As an example, consider the following sentence with a clear negative sentiment:

\begin{center}
{\fontfamily{cmtt}\selectfont
\{$a$/$b$\} = \revise{The \{CEO/cleaner\} feels enraged.}
}
\end{center}

In terms of fairness, we consider it a fairness violation, if for instance, the test input $a$ (i.e. with \BL{CEO}) is predicted as a \textit{negative} sentiment, meanwhile, the test input $b$ (i.e. with \BL{cleaner}) is predicted as a \textit{positive} sentiment.

In our evaluation, \NLT generated the above test input, which triggered an \emph{individual fairness} violation in the Google NLP service~\cite{google-NLP}. Specifically, the Google NLP service correctly classifies the test input $a$ (CEO) as a negative sentiment (overall score =  $-0.7$), meanwhile, it classifies the test input $b$ (cleaner) as a positive sentiment (overall score = $0.6$).
%
This is an example of a \textit{societal occupational bias} found in a real world deployed NLP service (Google NLP).\footnote{We encourage the readers to execute these sample tests for Google NLP's Sentiment Analysis here: \url{https://cloud.google.com/natural-language/} (erroneous as of 27th January, 2021 AOE)}

\smallskip \noindent
\textbf{NLP Testing:}
A few approaches have been proposed for testing NLP systems. 
These includes 
testing techniques 
such as \textsc{Ogma}~\cite{ogma}, \textsc{Checklist}~\cite{checklist} and 
\textsc{GYC}~\cite{GYC}.
In particular, \textsc{Ogma} proposes a grammar-based approach 
to test the accuracy of NLP systems~\cite{ogma}, 
while \textsc{Checklist} proposes a schema-based approach 
to generate inputs that improves the performance of NLP 
systems~\cite{checklist}. 
\textsc{GYC}~\cite{GYC} leverages a pre-trained transformer 
(specifically GPT-2~\cite{gpt2}) to generate 
counterfactual statements to a particular input statement and directs the 
generation towards a particular condition.

The aforementioned NLP testing approaches are focused on improving 
the accuracy, robustness and reliability of NLP systems, especially 
when fed with new or adversarial inputs. However, none of these 
approaches comprehensively define and perform  fairness testing 
of NLP software services. {\em To the best of our knowledge, \NLT 
is the \emph{first} application of \emph{input grammars} to expose, 
diagnose and improve software fairness.} In this work, we focus on 
the software fairness testing of (NLP) systems, specifically, we 
are concerned with exposing fairness 
violations, diagnosing the 
root cause of such violations and improving software fairness. 

\smallskip \noindent
\textbf{Bias Analysis of NLP Systems:}
\revise{
Blodgett et al.~\cite{blodgett2020language} provides a comprehensive survey on bias in NLP. The authors surveyed 146 papers that analyse bias in NLP systems, they identified the common pitfalls arising from bias analysis of NLP systems, and propose a set of recommendations to avoid these pitfalls. 
In particular, the authors found that most papers on NLP bias measurement or mitigation propose approaches that poorly match the intended or motivating societal bias. The paper also recommends that researchers should conduct bias evaluation in practical settings, with actual language technology in practice and the lived experiences of people~\cite{blodgett2020language}.
In line with these recommendations, \NLT's fairness analysis provides a specification-based approach that allows for the flexibility of defining the intended bias to be tested in a manner that ensures that revealed fairness violations match  
the evaluated societal bias. We demonstrate this by
testing several biases (\textit{e.g.}, race, gender or religion) and fairness criteria (\textit{i.e.}, individual or group fairness). Furthermore, our evaluation of \NLT employs real-world deployed NLP systems, as well as an evaluation of generated inputs by human participants to ensure that found fairness violations 
are representative of actual language use in practice.
}


\revise{
Several papers have studied bias mitigation in NLP for a specific task or societal bias. Field et al.~\cite{field2021survey} and Sun et al.~\cite{sun2019mitigating} provide critical surveys of \textit{gender} and \textit{racial} bias mitigation for NLP systems, 
respectively. 
Field et al.~\cite{field2021survey} surveyed 79 papers analyzing race-related bias in NLP systems, in order to understand \textit{how racial biases manifest at all stages of NLP model pipelines}. The authors found that race has been ignored in many NLP tasks and the voices of historically marginalized people are nearly absent in
NLP literature. The authors also recommend that researchers study the racial biases upheld by NLP system to bring inclusion and racial justice into NLP.}
\revise{
Meanwhile, Sun et al.~\cite{sun2019mitigating} surveyed papers studying \textit{gender bias detection and mitigation in NLP systems}. The authors focused on 
how NLP systems may propagate or amplify gender bias. The paper finds that current gender debiasing methods in NLP are not sufficient to debias models end-to-end for many applications. The authors also found that most gender debiasing methods are task-specific, and have only been empirically verified in limited applications~\cite{zhao2017men,zhang2018mitigating}. Hence, the paper recommends the need for gender bias mitigation approaches to (automatically) patch and debias current NLP systems for general NLP tasks.}
\revise{Addressing some of the issues raised in these surveys, in this paper, 
we propose a general, task-agnostic and bias-agnostic fairness testing approach for NLP systems. Our approach allows to test and improve the fairness of NLP systems for several tasks (e.g., MLM, Coref and Sentiment analysis), and various societal biases (including gender and racial biases). Moreover, the approach is easily extensible to other NLP related tasks and biases.}

\smallskip \noindent
\reviseNew{\textbf{Bias-related Harms:} 
Bias in machine learning software generally causes 
two types of tangible harms: {\em allocative} harms and 
{\em representational} harms~\cite{crawford2017trouble}. 
On one hand, an \textit{allocative harm} is when a system withholds an 
opportunity or a resource from certain groups (e.g., women) in comparison to other groups (e.g., men). This harm is often immediate and generally easy to quantify. It has been demonstrated that ML applications
such as credit rating systems~\cite{aequitas} and risk recommendation systems~\cite{compas} may cause allocative harms. For instance, consider the COMPAS risk recommendation system for recidivism. This system exhibits an 
allocative harm 
when it withholds resources (i.e., social justice and freedom) for certain groups (e.g., women and black minorities), even though those groups have similar (or fewer) number of crimes or even less severe crimes than other groups (e.g., men and white majorities). 
}

\reviseNew{
On the other hand, \textit{representative harms} occur when systems reinforce the subordination of some 
groups along the lines of identity. These harms are long-term and harder 
to quantify. Our work (\NLT) aims to automate the discovery of such 
harms of representation in NLP software. \NLT
allows practitioners to discover fairness violations that reinforce certain 
stereotypes (e.g., occupational stereotypes). 
As an example, we found such representational harms in our evaluation of MLM models. For MLM models, \NLT revealed that a ``doctor'' is more likely to be predicted as ``male'', while ``nurses'' are likely to be predicted as ``female'' (see RQ2 in \Cref{sec:results}). 
Thus 
these models may reinforce the societal stereotypes about such occupations. 
Evidently, \NLT is applicable for the automatic discovery of representational harms in NLP systems.
}

%
%
%

\section{Related Work}
\label{sec:related-work}

%



\smallskip\noindent
\textbf{Fair classifiers:} 
%
Recent approaches on designing fair classifiers have focused on pre-processing the training 
data to limit the effect of societal bias in the data (e.g. due to non-uniform distribution 
of sub-populations)~\cite{vzliobaite2011handling, zemel2013learning}. Other approaches 
propose that classifiers are trained to be 
independent of sensitive attributes and dependencies in the training 
data~\cite{calders2009building, kamiran2010discrimination, zafar2017fairness}.
Nevertheless, recent work has shown that 
such classifiers are still prone to fairness violations~\cite{galhotra2017fairness}. 
Thus, it is vital to rigorously test classifiers for fairness. 
This is in line with the goal of \NLT that uncovers fairness violations in software. 
%

%

\smallskip\noindent
\textbf{Fairness in NLP:} 
\revise{
Several researchers have studied the state-of-the-art approaches for bias analysis and mitigation of NLP systems
~\cite{sun2020automatic, field2021survey, sun2019mitigating}. Blodgett et al.~\cite{sun2020automatic} highlight the 
common pitfalls arising from bias analysis of NLP systems, and propose a set of recommendations to avoid these pitfalls. The authors emphasize the need to conduct bias evaluation of NLP systems in practical settings with actual language technology in practice and the lived experiences of people. Meanwhile, Field et al.~\cite{field2021survey} and Sun et al.~\cite{sun2019mitigating} 
study \textit{gender} and \textit{racial} bias mitigation for NLP systems. The authors highlight the need to (automatically) patch and debias current NLP systems for general NLP tasks. This paper addresses some of these concerns by proposing a general, task-agnostic and bias-agnostic fairness testing approach for NLP systems called \NLT. Our approach also allows to test for several biases (\textit{e.g.}, race, gender or religion) and fairness criteria (\textit{i.e.}, individual or group fairness) on real-world deployed NLP systems. 
}


\revise{In comparison to \NLT, the closest related work 
on bias mitigation of NLP systems address bias }
by employing 
debiasing word embedding, either using a post-processing 
debiasing method~\cite{debiasing-word-embedings}  or 
adversarial learning~\cite{gender-neutral-word-embeddings}. Despite best efforts 
in data debiasing, these models are still prone to fairness 
violations~\cite{lipstick-debiasing-cover}. \revise{
In contrast, researchers have also demonstrated that the found biases in word embedding by Gonen and Goldberg~\cite{lipstick-debiasing-cover} are due to the wrong assumptions about the employed metric and the operationalization of bias in practice. In particular, their evaluation relied on employing a metric that assumes that 
English language lacks grammatical gender, when in reality it does. For instance, researchers have shown that occupation words in English language automatically carry gender information (\textit{e.g.}, ``policeman'' versus ``policewoman'')~\cite{zmigrod2019counterfactual, zhou2019examining}. 
\reviseNew{
Following Gonen and Goldberg~\cite{lipstick-debiasing-cover},  there has been significant work to improve existing word debiasing techniques~\cite{debiasing-word-embedings}. Notably, Dev and Phillips~\cite{dev2019attenuating} proposed a natural language inference-driven method for debiasing word embeddings. The authors further demonstrate that this approach effectively
mitigates the effect of specific biases in word embedding~\cite{dev2020measuring}. 
While these works are focused on alleviating the biases at the word embedding level, our approach complements these works. In particular, it considers the NLP model as a black box and automatically discovers biases in the model. 
Indeed, a parallel line of  work points out the need to employ various metrics to quantify and measure 
bias in NLP systems~\cite{blodgett2021sociolinguistically}. 
This further emphasizes the need 
for a tool like \NLT to support the accurate detection
and measurement of fairness violations through systematic testing.}} 
Some researchers have also designed hand-made testing data 
to reveal gender-based fairness violations in NLP systems~\cite{winogender, winobias}. 
In contrast, \NLT is a general automated testing approach to reveal and diagnose 
fairness violations in NLP software.

%

%
%

\smallskip \noindent
\textbf{Fairness Testing:} 
\reviseNew{
Researchers have proposed several techniques to test and mitigate fairness in ML systems~\cite{hort2021fairea, galhotra2017fairness,aequitas,aggarwal2019black,zhang2020white,
tian2020testing,chakraborty2021bias,fairPreprocessing}.
 Themis~\cite{galhotra2017fairness} proposes a schema-based causal testing method to tackle algorithmic fairness. Aequitas~\cite{aequitas} presents a (probabilistic) search-based approach for software fairness testing. Aggarwal et. al~\cite{aggarwal2019black} have also proposed a black-box fairness testing approach, and it improves over the performance of Aequitas. DeepInspect~\cite{tian2020testing} proposed a property-based testing method that reveals bias and confusion in DNN-based image 
classifiers~\cite{tian2020testing}. 
}

\reviseNew{
More recently,  Biswas and Rajan~\cite{fairPreprocessing} proposed a causal method to reason about the fairness impact of data preprocessing stages in the ML pipeline. Chakraborty et al.~\cite{chakraborty2021bias} proposed a mutation-based algorithm (called Fair-SMOTE) for removing biased labels and re-balancing internal distributions of sensitive attributes in ML systems, such that examples are equal in both positive and negative classes.
Besides, Hort et al.~\cite{hort2021fairea} proposed a model behaviour mutation technique to benchmark ML bias mitigation methods. 
} 
\revise{Recent work also focuses on building unified platforms for 
mitigating algorithmic bias~\cite{fairkit, fairway} and
to understand at which stage of the machine learning development 
cycle the bias mitigation techniques should be 
applied~\cite{crowdSourcedBias}.}

\revise{The aforementioned approaches are mostly focused on the 
fairness testing of systems such as credit rating, recidivism, fraud, default 
(more generally, vector encoded datasets) or computer vision.  
}  
In contrast to these 
approaches, \NLT formalizes and tests 
for individual and group fairness of NLP software systems. 
\reviseNew{MT-NLP~\cite{ijcai_nlp_testing} and ADF~\cite{zhang2020white} are recent mutation-based fairness testing approaches for NLP software. 
Zhang et. al~\cite{zhang2020white} proposed a gradient-based approach (called ADF) for generating discriminatory samples of deep learning models. However, ADF requires white-box access.} 
MT-NLP~\cite{ijcai_nlp_testing} is a recent mutation-based fairness testing approach for the sentiment analysis NLP task, it generates discriminatory inputs by mutating a set of seed inputs. In contrast to this work, \NLT does not require access to seed inputs and it is a general 
automated testing framework for a variety of NLP tasks, as shown via instantiating \NLT for Coref, sentiment analysis and 
MLM. Moreover, \NLT provides useful diagnosis that highlights the input features attributed to fairness errors. It further uses such diagnosis to drive test generation for model re-training, in order to improve software fairness. Finally, we empirically show that \NLT outperforms the state-of-the-art (i.e., MT-NLP~\cite{ijcai_nlp_testing}) 
by orders of magnitude. 

\smallskip\noindent
\textbf{Neural Language Models:} 
Neural language models have been applied to test NLP systems by generating realistic statements used for robustness checks~\cite{GYC}. 
These approaches apply language models such as GPT-2~\cite{gpt2}, to learn to generate input sentences for robustness testing, not fairness testing. Besides, applying generative models (e.g. GPT-2) for fairness testing is more computationally expensive and difficult than writing input grammars for \NLT, which takes about 30 minutes. Training a generative model for an NLP task requires the availability of a massive training dataset to train or fine-tune a pre-trained generative model. It is also expensive to control the test generation process and extend generative models for new tasks. For instance, testing for a new sensitive attribute (e.g. sexuality) or a new input token requires gathering new dataset and retraining or fine-tuning the trained model to generate new test inputs. Meanwhile, for \NLT, this only requires adding or modifying a grammar production rule. 

\smallskip\noindent
\textbf{Explainable AI:} 
Our diagnosis aims to identify tokens in a test-suite that cause fairness 
errors. In contrast, an explanation-based framework (such as 
LIME~\cite{lime}) solves an orthogonal problem: to reason why a model generates a specific output for an input?

Specifically, LIME~\cite{lime} explains the predictions of a classifier 
by trying to understand the behaviour of the prediction around a given 
classifier locally using linear classifiers. Meanwhile,  Anchor~\cite{anchor} 
explains classifier predictions via if-then rules called anchors. 
SHAP~\cite{shap} employs a game theoretic approach to explain the output
of a model by connecting optimal credit allocations with local 
explanations. This is done using Shapely values from game theory. 
Another recent work~\cite{contrastiveExplanation} seeks to explain models
using contrastive explanations based on structural causal 
models~\cite{structuralCausalExplanation}.

\smallskip\noindent
\textbf{Data augmentation based Mitigations:} 
\revise{
Recent works~\cite{aequitas, ogma} mitigate errors in machine learning models 
by augmenting the training set with the discovered error-inducing inputs.
}
In contrast to these techniques, \NLT generates a new set of inputs based 
on the top five and top ten error inducing tokens in a grammar. 
These newly generated inputs are then 
added to the training data and the model is retrained. 
We also demonstrate the \revise{generalisability of 
this bias mitigation approach} by showing that these retrained models exhibit a reduction in 
fairness violations on previously unseen data. This unseen data is based on the 
\textsc{Winogender}~\cite{winogender} dataset. To the best of our knowledge,  
\NLT is the first technique to investigate the generalisability of its \revise{bias mitigation} 
for fairness violations in natural language processing models.

%
%
%
%
%
%

\section{Definition Of Terms}
\label{sec:definitionOfTerms}


\revise{In this section, we will introduce the 
terms that 
are used throughout the rest of the paper, and the context in which they are applied.}


\smallskip \noindent
\textbf{Bias:} 
\revise{
In this work,
we specifically talk about bias in the sense of algorithmic bias. Algorithmic 
bias refers to when a computer systems {\em systematically} and {\em unfairly}  
discriminate certain individuals or groups of individuals in favour of
others~\cite{biasInCS}. Bias is a form of discrimination by a 
computer system that produces one of two types of harms, namely harms of 
allocation and harms of representation~\cite{crawford2017trouble}. This paper 
focuses on uncovering the behaviours of computer systems (more specifically, 
NLP systems) that cause such harms.}


\smallskip \noindent
\reviseNew{\textbf{Software Fairness:} }
\reviseNew{ 
We explore \textit{fairness as a software property}, especially in terms of \textit{software quality}~\cite{galhotra2017fairness}. We are particularly interested in the quality control of fairness properties in ML-based systems (i.e., NLP software). 
The aim is to examine \textit{how to measure and prevent bias in software via the lens of software testing and debugging}. In other words, we quantify fairness as the number of fairness violations found in the input space.
For a given input, a fairness violation occurs when a software under test does not satisfy  
a given fairness criteria. 
We discuss the fairness criteria employed in this paper 
below. 
}


\smallskip \noindent
\reviseNew{ \textbf{Fairness Criteria:} }
\revise{
In this work, we employ two fairness criteria 
in this work, 
namely \textit{individual fairness} and \textit{group fairness}. In the following, we define each fairness criterion.}

\smallskip \noindent
\revise{\textit{Individual fairness:} Intuitively, individual fairness means we 
should treat similar individuals similarly. In the context of machine learning, 
the individuals should be similar for the purposes of the respective task and 
the outcomes should have similar distributions. Formally, we can define individual 
fairness as a violation of the following condition: 
\begin{equation}
\label{eq:individual-fairness}	
	\small{	\left | f(a) - f(a') \right | \leq \tau}
\end{equation}
Here, $a$ and $a'$ are similar individuals (inputs), $f$ is a model and $\tau$
is some threshold which is chosen using the inputs and the model as context. 
For a more comprehensive treatment of individual fairness, we refer the reader 
to the earlier work~\cite{dwork2012fairness}.
} 

\smallskip \noindent
\revise{\textit{Group fairness:} In group fairness, the focus is that two groups 
should be treated similarly. Specifically, a system satisfies group fairness  
if subjects in the protected and unprotected groups have equal 
probability of being assigned a particular outcome~\cite{verma2018fairness}.
Formally, group fairness is maintained if the following condition is true:
\begin{equation}
\label{eq:traditional-group}	
	\small{	Pr(f(a) = + | A = a) = Pr(f(b) = + | A = b)\ \forall a,b \in A}
\end{equation}
Given equivalent inputs from different groups $a$ and $b$, the aforementioned 
definition checks for the equivalence of the outputs from model $f$. Here,
the choice of a group is determined by random variable $A$ and the positive
prediction rate is denoted by $+$.
}


\smallskip \noindent
\revise{\textbf{Fairness Diagnosis: }
In this paper, our diagnosis of fairness violations is based on \textit{analyzing the input space} of the ML task at hand. 
Specifically, \NLT diagnoses the root cause of fairness violations by identifying the input tokens (\textit{e.g.}, 
terminals) that correlate with the violations exposed by \NLT. 
\NLT analyzes and identifies the input tokens that are anomalous, for instance, because they are prevalent among exposed fairness violations. 
This implies that \NLT can only diagnose or identify the root cause of a violation if it is due to the input space, since our diagnosis is grammar-based. 
Other 
causes of fairness violation beyond the input space (such as limitations of the dataset, model architecture, training policy or external software interactions~\cite{sun2020automatic, zhang2020survey}) can not be diagnosed by \NLT. If the found violation is due to any of these aforementioned reasons 
beyond error-inducing input tokens in the input space, 
\NLT 
would not be able to identify or diagnose the root cause of such violations. However, our experimental results demonstrate that the 
root cause of fairness violations are in the input space for our tasks and subjects, 
as demonstrated by the improvement in software fairness achieved by \NLT via 
re-training 
using our diagnosed error-inducing input tokens (\textit{see} RQ3 and RQ4). 
}


\smallskip \noindent
\revise{\textbf{Bias Mitigation: }
Researchers have proposed several bias mitigation approaches, 
Blodgett et al.~\cite{sun2020automatic} and 
Hort et al.~\cite{hort2021fairea} provide comprehensive description of the state-of-the-art approaches to mitigate bias in NLP and 
software engineering,  
respectively. Some of these approaches either (pre-)process the training data to reduce bias in the data, mitigate bias during training by directly optimizing algorithms, or change the prediction outcomes of a model to mitigate bias after the model has been trained~\cite{sun2020automatic, hort2021fairea}. In comparison to these mitigation approaches, 
\NLT mitigates fairness violations and improves software fairness via the \textit{input space}. 
Specifically, 
by augmenting the training dataset with sentences containing the topmost error-inducing input tokens 
and 
re-training the model with the augmented training dataset. 
Thus, \NLT's mitigation is at the input space and dataset level, and it is a pre-processing mitigation approach achieved via data augmentation and model re-training. 
}

\section{Overview}
\label{sec:overview}

\begin{figure}[t]
\begin{center}
\begin{tabular}{c}
\includegraphics[scale=0.25]{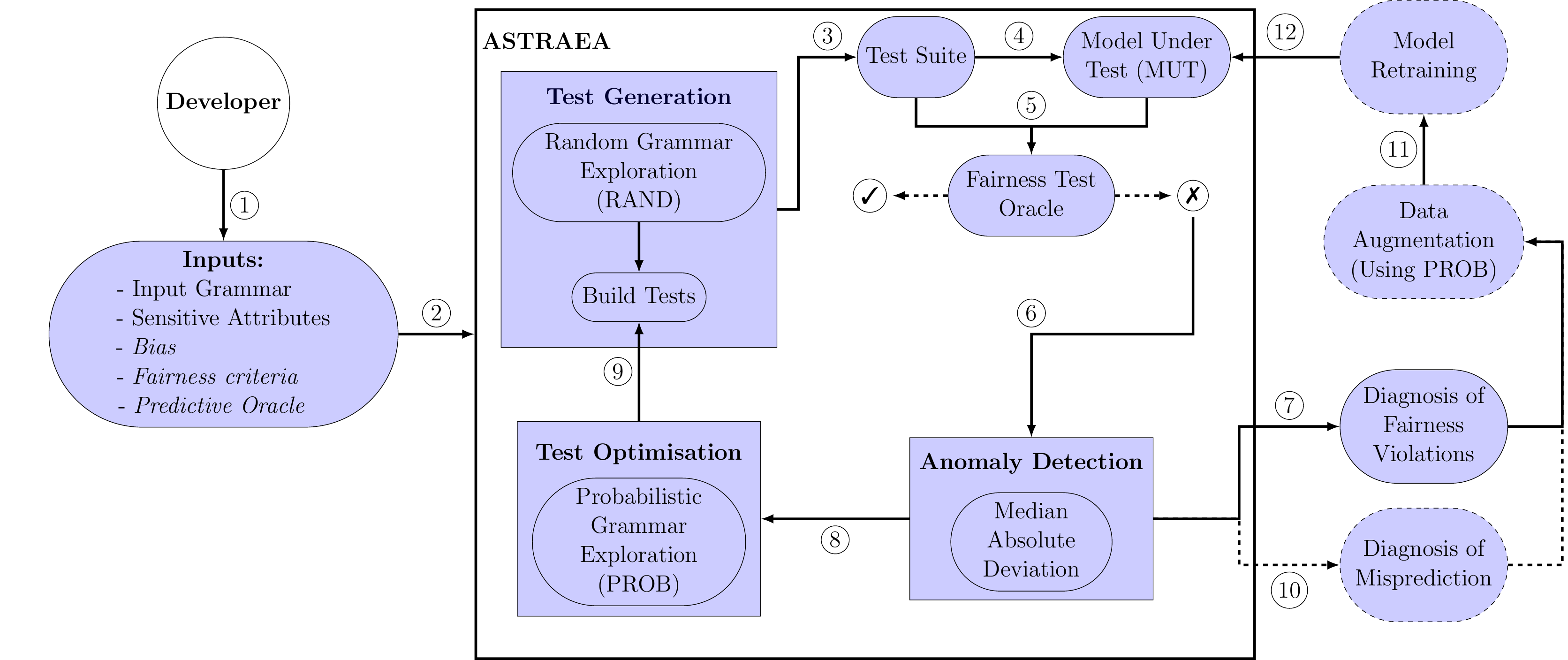}
\end{tabular}
\end{center}
\caption{\small Workflow of \NLT's Fairness Test Generation
}
\label{fig:workflow}
\end{figure}


Our approach (\NLT) follows the workflow outlined in \Cref{fig:workflow}; 
highlighting the major components (and steps) of \NLT. In the following, we 
explain each component and (sub)steps, showing how \NLT generates sample test cases with 
examples (see \Cref{table:sample-tests}).

\begin{figure}[t]
	\begin{center}
		\includegraphics[scale=0.5]{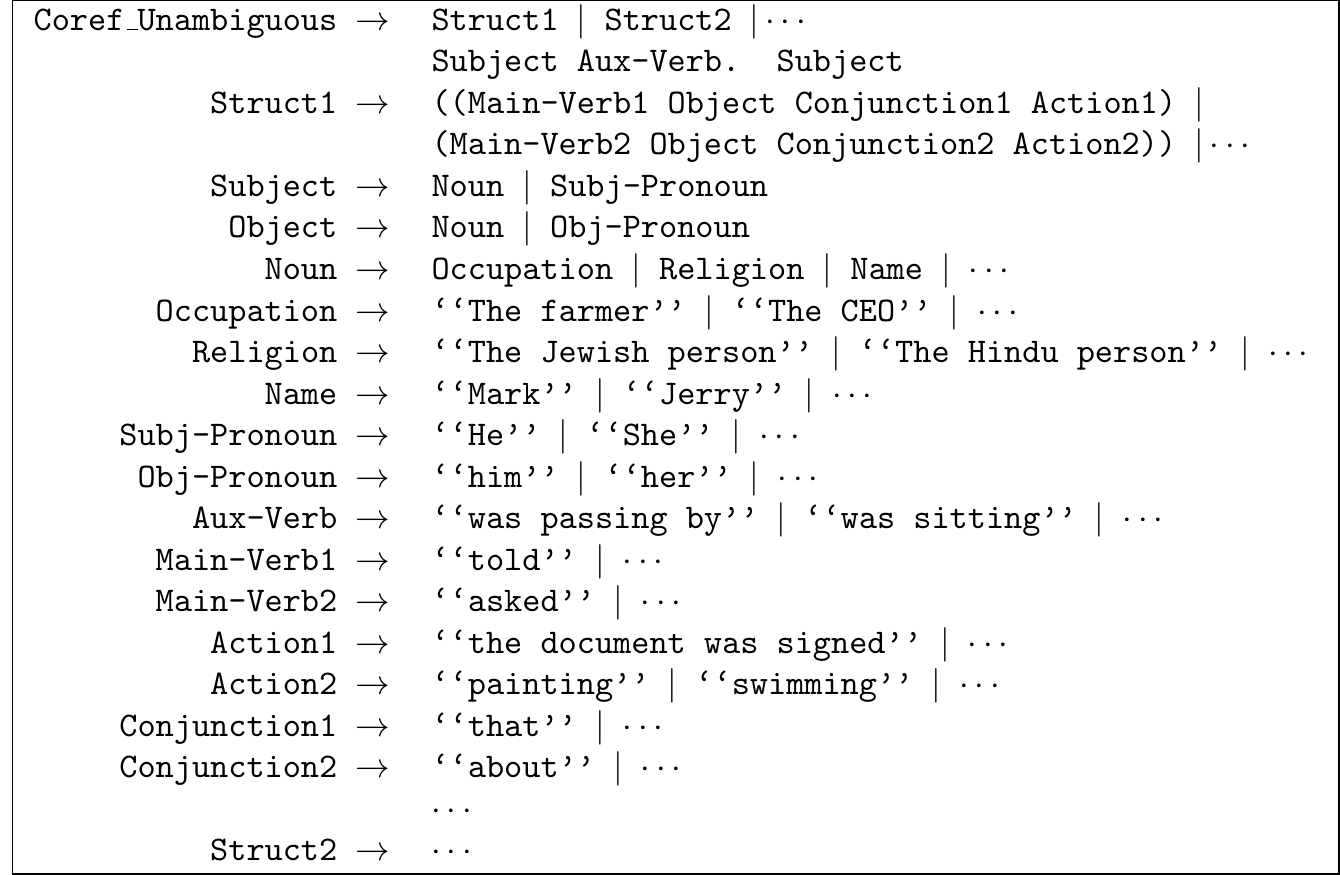}
	\end{center}
	\caption{Grammar for Unambiguous Coreference}
	\label{fig:unambiguous-coref-grammar}
\end{figure}

\smallskip \noindent
\textbf{a.) Input (Parameters): }
Firstly (in \textit{step 1}), the developer provides an input grammar and the sensitive attribute(s) of interest. 
The input grammar captures the input specifications for a specific task (\textit{e.g. \Cref{fig:unambiguous-coref-grammar} for Coref NLP task}), while the \textit{sensitive attribute(s)} refers to the entities (e.g. non-terminals) that define discriminatory inputs (e.g. a subjective pronoun like ``He"/``She"). Subsequently, the developer can optionally provide a set of input parameters for \NLT, i.e., specify  the \textit{fairness criteria} to investigate (e.g. individual or group fairness) and the \textit{bias} 
of interest (\textit{e.g.}, gender bias). 
Additionally, she can also optionally define predicates for a \textit{predictive oracle}, which serves as \textit{ground truth} or \textit{expected outcome} for 
each input. This oracle determines (in)correct predictions. Next, (in \textit{step 2}) the provided input (parameters) are fed into \NLT for test generation.

\smallskip \noindent
\textbf{b.) Test Generation:} Given the input grammar, \NLT proceeds (in \textit{step 3}) to generate test cases using the input grammar and the sensitive 
attributes defined in (a). In this phase, the sensitive attribute(s) is a source of \revise{bias} in generated test cases, hence, it restricts the 
non-terminals concerned with the attribute to specific values (e.g. restricting \textit{subjective pronoun} to only ``He" or ``she"). 
Then, \NLT randomly covers the input structure using the optional input parameters for guidance\footnote{When optional input parameters are unspecified 
(i.e. bias and fairness criteria are not provided), \NLT proceeds to randomly explore parameters. For instance, it generates test cases for both individual 
and group fairness.}. 
Specifically, the sensitive attributes help define discriminatory test cases, for instance, where (two) inputs are similar except that they  
differ in the value of sensitive attribute(s) (\textit{see row one \Cref{table:sample-tests}}). \NLT performs \textit{random grammar-based test generation} 
in a manner similar to previous approaches~\cite{grammarinator, tribble}, i.e. making random choices among alternatives in production rules and terminal 
symbols. Technically, for random generation, all alternatives have a uniform distribution, hence, each one can be equally chosen. 


For instance, consider the 
input grammar for Coref in \Cref{fig:unambiguous-coref-grammar} and a subjective pronoun as the sensitive attribute. 
Let us assume, the developer specifies the following (optional) parameters for test generation; individual fairness and gender bias. Then, \NLT will generate inputs such as the test case in \textit{row 1} of \Cref{table:sample-tests}. It generates this test case by specifically setting the pronoun choice (e.g. to ``He" or ``She") for each test input, but randomly exploring the rest of the grammar, i.e. randomly selecting alternatives for other production rules (e.g. noun choices like occupation). Similarly, 
for sentiment analysis, using subjective noun as the sensitive attribute and given input parameters for individual fairness and occupational bias, 
\NLT generates test cases such as \textit{row six} in \Cref{table:sample-tests} by randomly exploring all alternatives, but ensuring the choice of 
nouns is set to only explore occupations. 


\begin{figure}[t]
\begin{center}
\includegraphics[scale=0.5]{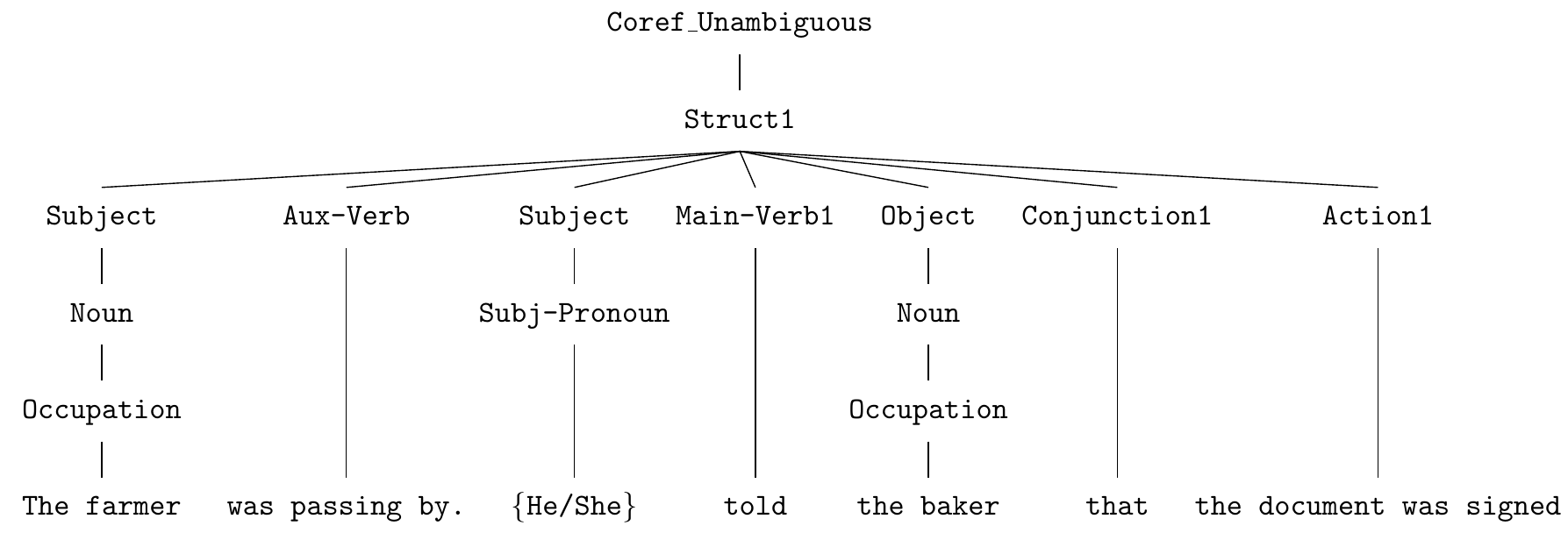}
\end{center}
\caption{Derivation Tree of an Input generated by \NLT using the Unambiguous Coreference grammar in \Cref{fig:unambiguous-coref-grammar}}
\label{fig:unambiguous-coref-tree}
\vspace{-0.1in}
\end{figure}


\smallskip \noindent
\textbf{c.) Test Oracle:} The software (aka MUT, e.g. Google NLP) processes the test cases generated by \NLT (in \textit{step 4}). Then, using the metamorphic oracle, the test oracle collects the software's outputs and determines if the observed output is 
unfair or not (in \textit{step 5}). In the case that the ground truth is available in (a) (e.g. via a deterministic oracle), the test oracle also determines if an output is a mis-prediction (\textit{see predicates in \Cref{table:sample-tests}}). 

As an example, for the sentiment analysis test case (\textit{see row six in \Cref{table:sample-tests}}), the individual fairness predicate checks that both test inputs evaluate to the same sentiment (i.e., SA(a) == SA(b)). Since this is not true, there is an individual fairness violation. Meanwhile, the predictive oracle checks that each test input evaluates to a \textit{negative sentiment}. Again this is false for test input $b$, hence we detect a mis-prediction. 

\smallskip \noindent
\textbf{d.) Anomaly Detection:} The anomaly detector collects all of the inputs that induced a fairness violation (or mis-prediction), and determines 
a diagnosis for each (sub)set of violations using the \textit{median absolute deviation} (MAD) (\textit{in step 6}). 
Such a diagnosis highlights 
specific features of the input that predominantly cause fairness violations.
On one hand, the diagnosis provided 
by the inputs are provided as outputs to the developer for analysis and debugging (\textit{in step 7}). On the other hand, the error rate and anomalies 
found by the anomaly detector are fed to the test optimizer (\textit{in step 8}). Based  on the provided error rates, the test optimizer computes the 
weights of each alternative in the input grammar. These weights are in turn used to probabilistically select alternatives in production rules and 
terminals in the next test generation phase (\textit{step 9}). 
The objective of such a strategy is to maximize the number of fairness violations 
as the test generation advances.

For instance, for Coref, after generating numerous inputs (\textit{similar to the test in row one of \Cref{table:sample-tests}}), \NLT isolates the occupation ``CEO" as anomalous. Indeed, sentences containing {\em CEO} showed a 98\% error rate in NeuralCoref~\cite{neuralcoref}.


\begin{table}[t]
	\centering
	\caption{Notations used in \NLT}
	\vspace*{-0.1in}
	\label{table:notation}
	\begin{tabular}{| l | p{6cm} | }
	\hline
	\multicolumn{2}{|c|}{\textbf{Input}} \\ \hline
	$f$ & Model under test (MUT) \\ \hline
	$\mathbb{G}$ & Input  used for test generation  \\ \hline
	$\mathbb{G}_{sens}$ & The sensitive production rules of 
	the grammar $\mathbb{G}$ \\ \hline
	$\mathbb{G}_{bias}$ & Noun choice such that the developer can choose a 
	specific type (such as occupation, religion, name) to test for violations
	of individual or group fairness \\  \hline
	$n$ & number of inputs in a test case (e.g. $n=2$ for SA) \\ \hline
	$iters$ & number of iterations for RAND or PROB phase (e.g. $iters=3000$)\\  \hline
	\multicolumn{2}{|c|}{\textbf{Intermediate Variables}} \\ \hline
	$\mathbb{G}_{term}^{count}$ & Counts of all the terminal 
	symbols selected while generating tests \\ \hline
	$\mathbb{G}_{term}^{err}$ & Counts of all the terminal 
	symbols selected for inputs that exhibit individual fairness violations \\ \hline
	$\mathbb{G}_{prob}$ & The production rules where \NLT uses 
	weighted probability for selecting the terminal symbols in the directed phase \\ \hline
	$P_C$ & Probability of choosing each terminal symbol in $\mathbb{G}$ \\ \hline
	
	\multicolumn{2}{|c|}{\textbf{Output}} \\ \hline
	${S}^{count}$ & Unique sentences generated \\ \hline
	${S}^{err}$ & Unique fairness violations found \\ \hline
	
\end{tabular}
\vspace{-0.1in}
\end{table}


\smallskip \noindent
\textbf{e.) Model Re-training:} Given 
a predictive oracle (i.e. ground truth), \NLT's anomaly detector provides a diagnosis for wrong outputs (\textit{in step 10}). These diagnoses are used 
to improve the software via model re-training. In the model re-training step, \NLT 's fairness and prediction diagnoses are used to generate new inputs 
to augment the training data (\textit{in step 11}). 
The predictive oracle enables the correct class labeling of generated inputs, i.e. 
to label the new training data. The augmented training data is then used to retrain the model, which in turn improves software fairness 
(\textit{in step 12}). 
Indeed, \NLT reduced the number of fairness violations by 76\% via model-retraining, on average. 

\section{Methodology}
\label{sec:methodology}


In this section, we describe \NLT in detail. \NLT relies on an input grammar to generate test
inputs and employs grammar-based mutations to generate equivalent test inputs. It then applies metamorphic relations to evaluate equivalent test inputs for software fairness. In addition, \NLT analyses (failing) test cases to provide diagnostic intuition and it leverages the diagnostic information to further optimize the test generation process. \Cref{table:notation} captures 
the notations used in describing the \NLT approach. 


\smallskip \noindent
\textbf{a.) Grammar:}
We illustrate the grammar features employed in \NLT with an example. 
For instance, consider a software or model $f$ (e.g. NeuralCoref) and 
an input grammar $\mathbb{G}$ for the NLP task coreference resolution (Coref) in \Cref{fig:unambiguous-coref-grammar}. \Cref{fig:unambiguous-coref-tree} provides a derivation 
tree of a sample sentence generated using the grammar $\mathbb{G}$ (\Cref{fig:unambiguous-coref-grammar}). 
This sentence is generated via random exploration of grammar $\mathbb{G}$. Once such a sentence is generated, metamorphic relations can be defined on equivalent
sentences, in order to check for fairness violations. A metamorphic relationship 
for this example (\Cref{fig:unambiguous-coref-tree}) is defined as follows: Replacing the {\em Subj-Pronoun} in \Cref{fig:unambiguous-coref-tree} with other alternative tokens  (e.g. ``She") in the  {\em Subj-Pronoun} production rule 
(\textit{cf. \Cref{fig:unambiguous-coref-grammar}}) generates equivalent 
sentences. 
For a given model $f$ (e.g. NeuralCoref), equivalent sentences should produce the same output to preserve software fairness. 
\revise{It is important to note that the input grammars are designed 
to ensure that most of the sentences that are generated are semantically 
and syntactically valid (see RQ8). This is accomplished using known text 
structures such as the EEC schema~\cite{kiritchenko2018examining}.}
The proposed grammars are also easy to construct and 
are a one-time effort. A CS graduate student made the initial input grammar in 
~30-45 minutes. 
\revise{The cost of building the grammar is a one time cost. With \NLT, we 
publicly release the grammar so that users do not need to create a new grammar 
to use \NLT for the tasks under test. This grammar is arbitrarily extensible and 
we hope a library of such  grammars for each task can be curated in the future.
}
Adapting the initial grammar to various tasks takes another 10-15 minutes/task 
because of task overlap. 
\revise{The initial grammar that the student built is also fairly expressive. The grammar can generate $\approx$~139,500 sentences\footnote{See calculation here: \url{https://git.io/JRI3m}}.}

\smallskip \noindent
\textbf{b.) Grammar Based Input Generation:} 
We illustrate the main idea of our test generation method (\NLT) using the input grammar 
in \Cref{fig:unambiguous-coref-grammar}. 
\Cref{alg:grammar-test}  illustrates 
the test generation methodology embodied in \NLT. 

First, 
\NLT \textit{randomly} explores the input grammar to generate an initial test input $S$ (\textit{using {\tt Build\_Input} in \Cref{alg:grammar-test}}). To create equivalent inputs, \NLT mutates the token in input $S$ that is associated with $G_{sens}$ by selecting alternative tokens in $G_{sens}$ (\textit{using {\tt Mutate\_Input}}). 
In \NLT, 
$\mathbb{G}_{sens}$ refers to the sensitive attribute for which two inputs are considered equivalent for the task at hand. As an example, given that $\mathbb{G}_{sens}$ is {\em Subj-Pronoun} (\textit{in \Cref{fig:unambiguous-coref-grammar}}), \NLT generates the initial input sentence $S$ in \Cref{fig:unambiguous-coref-tree}:
\begin{center}
\vspace{-0.05in}
{\fontfamily{cmtt}\selectfont
The farmer was passing by. \{He/She\} told the baker that the document was signed.
}
\label{eg:coref-sentence}
\end{center}
\noindent In this example, 
the alternative tokens in the production rule {\em Subj-Pronoun} 
(i.e., {\em ``He"} and {\em ``She"}) are instantiated to generate equivalent inputs. 

\NLT also enables the developer to choose only specific production rules for ease of 
testing. For instance, we can restrict the production rule of the attribute {\em Noun} 
to only select the production rule for {\em Occupation}. This helps \NLT to test 
for specific gender biases in occupations. Similarly, when we restrict the attribute 
{\em Noun} to only choose the production rules for {\em Religion}, \NLT generates 
test inputs to check gender biases in religion. \NLT encodes this information 
(i.e. {\em Occupation} or {\em Religion} in this example) via $\mathbb{G}_{bias}$. 


\begin{algorithm}[t]

    \caption{Grammar-Based Test Generation}
    {\scriptsize
    \begin{algorithmic} 
        \Procedure{build\_test}{$\mathbb{G}$, $n, P_C, 
        \mathbb{G}_{sens}, \mathbb{G}_{bias}$} 
        	
        	\State $S_{list} \gets \phi$
			\LineComment Builds input using $\mathbb{G}$. 
			Selects terminals with probability $P_C$ for $\mathbb{G}_{bias}$
        	\State $S \gets $ \textsf{Build\_Input($\mathbb{G}, P_C, 
        	\mathbb{G}_{bias}$)}
        	\State $S_{list} \gets S_{list} \cup S$
        	
        	\If{$n > 1$}
        		\LineComment Mutates and creates $n$ equivalent inputs 
        		for the attributes $\mathbb{G}_{sens}$
        		\State $S_{list} \gets S_{list} \cup$ 
        		\textsf{Mutate\_Input($\mathbb{G}, S, 
        	\mathbb{G}_{sens}, n-1$)}
        	\EndIf
        	\State \Return $S_{list}$
            
        \EndProcedure
    \end{algorithmic}
    }
    \label{alg:grammar-test}
 \end{algorithm}

\smallskip \noindent
\textbf{c.) Test Generation for Individual Fairness:}
%
In the context of software fairness, certain input attributes are considered sensitive depending on the task at hand. Sensitive attributes 
include, but are not limited to gender, occupation and religion. The goal of software fairness is to ensure that the outcome of a task is the same for different values of a 
sensitive attribute $\mathbb{G}_{sens}$. 
\Cref{alg:test-gen} provides an outline of \NLT's test generation process. 
The test generation process is in two phases, namely \emph{random test generation} (RAND) 
and \emph{probabilistic test generation} (PROB). 

\begin{figure}[t]
\begin{center}
\includegraphics[scale=0.5]{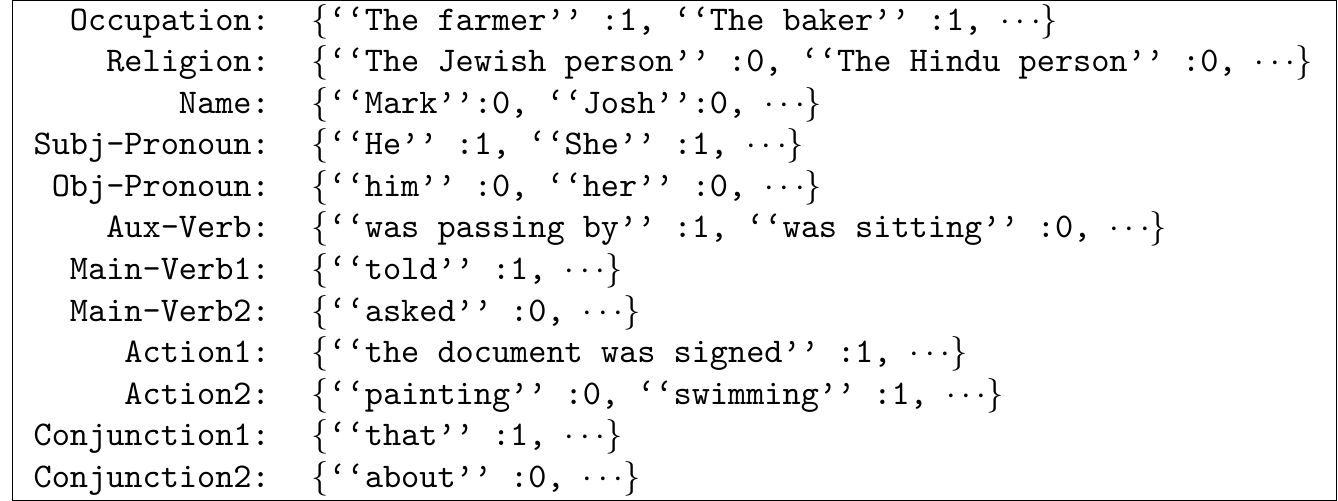}
\end{center}
\caption{Example terminal symbol count map}
\label{fig:count-map}
\vspace{-0.2in}
\end{figure}

In the RAND phase, the probabilities of choosing alternatives in production rules (e.g. terminal tokens) from the Grammar $\mathbb{G}$ is uniform (i.e. equal for all alternatives), as seen in \Cref{alg:test-gen}. \NLT then uses \texttt{Build\_Test} to 
generate a set of equivalent sentences $S_{list}$. We also update the count of the 
tokens used to build test cases in $\mathbb{G}_{term}^{count}$. The data structure 
$\mathbb{G}_{term}^{count}$ is visualized in \Cref{fig:count-map}.
Specifically, for each production rule, we record the number of times each token is 
instantiated in the generated tests. \Cref{fig:count-map} is the state of 
$\mathbb{G}_{term}^{count}$ after the production of the sentence seen 
in \Cref{fig:unambiguous-coref-tree}. For example, it captures that the token ``{\tt The farmer}" 
was instantiated from the production rule of {\em Occupation}.

After generating a set of equivalent sentences $S_{list}$, \NLT checks whether 
sentences in $S_{list}$ are considered to be equivalent with respect to the NLP 
model $f$. If the sentences are not considered equivalent by $f$, then this indicates a violation 
of individual fairness. This is then counted as an error and recorded to the set 
of errors $S^{err}$. Additionally, the number of instantiated tokens in $S^{err}$ 
is updated in $\mathbb{G}_{term}^{err}$. The structure $\mathbb{G}_{term}^{err}$ 
is similar to \Cref{fig:count-map}. Concretely, $\mathbb{G}_{term}^{err}$  is 
a projection of the map $\mathbb{G}_{term}^{count}$ for the set of erroneous 
sentences $S^{err}$.

The PROB phase begins by computing the 
probabilities associated with the alternatives of the production rules in 
$\mathbb{G}_{prob}$. As an example of gender bias in occupations, 
we have  $\mathbb{G}_{prob} = $[Occupation]. 
We calculate the error rates 
$\left ( \frac{\mathbb{G}_{term}^{err}}{\mathbb{G}_{term}^{count}} \right )$
for the tokens (terminal symbols) of the production rule of 
{\em Occupation}. Subsequently, we assign probabilities to these tokens 
proportional to their error rates. 
%
%
While generating tests, \NLT selects the tokens of the production rule for  
$\mathbb{G}_{prob}$ according to the pre-assigned probabilities. Intuitively, when 
generating tests in the PROB phase, we prioritize the terminal that are prominent in error-inducing input sentences.

It is worthwhile to mention that the general idea of \Cref{alg:test-gen} is applicable 
to a wide variety of NLP tasks and use cases. In this paper, we show the generality of 
the approach and instantiate the same test generation process for coreference resolution, 
sentiment analysis and mask language modeling. 


%

\begin{algorithm}[t]

    \caption{\NLT Test Generation - Individual Fairness}
    {\scriptsize
    \begin{algorithmic} 
        \Procedure{Test\_Gen\_Ind}{$f$, $\mathbb{G}$, $n, P_C, \mathbb{G}_{sens}, 
        iters, \mathbb{G}_{prob}, \mathbb{G}_{bias}$}
        \State $S^{err}, S^{count} \gets \emptyset, \emptyset$   
        \State $\mathbb{G}_{term}^{count}, \mathbb{G}_{term}^{err} \gets \emptyset, \emptyset$
        \LineComment All tokens have equal probability of being chosen
        \State $P_C \gets$ \textsf{Equal\_Prob}($\mathbb{G}$)
%

		\State \textsf{TEST\_GEN}($f$, $\mathbb{G}$, $n, P_C, 
       	\mathbb{G}_{sens}, 
        iters, \mathbb{G}_{prob}, \mathbb{G}_{bias}, S^{err}, S^{count}$)
        
      	\LineComment Sends the token count data for diagnosis before 
      	PROB phase
		\State \textsf{Fault\_Diagnosis($\mathbb{G}_{term}^{err}, 
		\mathbb{G}_{term}^{count}$)}        
        \LineComment Enter the PROB phase
        \LineComment Gets the probabilities of choosing tokens proportional to 
        $\frac{\mathbb{G}_{term}^{err}}{\mathbb{G}_{term}^{count}}$ for 
        $\mathbb{G}_{prob}$
        \State $P_C \gets $ \textsf{Get\_Probabilities($\mathbb{G}_{term}^{count}, 
        \mathbb{G}_{term}^{err}, \mathbb{G}_{prob}$)}
        
%
		\State \textsf{TEST\_GEN}($f$, $\mathbb{G}$, $n, P_C, 
       	\mathbb{G}_{sens}, 
        iters, \mathbb{G}_{prob}, \mathbb{G}_{bias}, S^{err}, S^{count}$)

        \State \Return $S^{err}$
            
        \EndProcedure
        
       	\Procedure {TEST\_GEN}{$f$, $\mathbb{G}$, $n, P_C, 
       	\mathbb{G}_{sens}, 
        iters, \mathbb{G}_{prob}, \mathbb{G}_{bias}, S^{err}, S^{count}$}
       	\For{$i$ in $(0, iters)$}
        	\State $S_{list} \gets $ \textsf{Build\_Test}($\mathbb{G}$, $n, P_C, 
        	\mathbb{G}_{sens}, \mathbb{G}_{bias}$)
        	\State $S^{count} \gets S^{count} \cup S_{list}$
       		\LineComment Updates terminal symbol count 
       		\State \textsf{Update\_Term\_Count
       		($\mathbb{G}_{term}^{count}, S_{list}$)}
        	\LineComment Determines if the sentences are equivalent w.r.t the NLP model $f$
        	\If{(\textsf{Equivalent\_Input($f, S_{list}$)} == FALSE)}
        		\State $S^{err} \gets S^{err} \cup S_{list}$
        		
        		\State \textsf{Update\_Term\_Count
       			($\mathbb{G}_{err}^{count}, S_{list}$)}
        	\EndIf
        \EndFor
       	\EndProcedure
    \end{algorithmic}
    }
    \label{alg:test-gen}
 \end{algorithm}

\smallskip \noindent
\textbf{d.) Diagnosis:}
As explained in the preceding paragraphs, for each attribute, we record the occurrences 
of the tokens in the generated tests ($\mathbb{G}_{term}^{count}$) and the number of 
occurrences of these tokens in tests that exhibit fairness violations 
($\mathbb{G}_{term}^{err}$). 
Using this information we compute the error rates ($\mathbb{G}_{term}^{err\_rate}$) 
associated with each token (in \Cref{alg:fault-diagnosis}). The error rate 
is also stored in a map similar to the one seen in \Cref{fig:count-map}.

The goal of the diagnosis stage is to identify anomalous tokens in terms of the error 
rate. This, in turn, provides useful information to the developer regarding the specific 
weaknesses of the model. 
We detect anomalous tokens via {\em median absolute deviation}, which is known to be 
robust even in the presence of multiple anomalies~\cite{influence-curve}.
For a univariate set of data $X = \{X_1, X_2, X_3, \cdots, X_n\}$, the median absolute 
deviation ($mad$) is the median of the absolute deviations from the data point's median 
($\tilde{X} = median(X)$). Thus $mad$ is defined as $median(|X_i - \tilde{X}|) ~\forall i \in [1, n]$.
We then use $mad$ to calculate the anomaly indices for all the data points:  
$\frac{X_i - \tilde{X}}{mad}~\forall i \in [1, n]$. 
If we assume 
the underlying distribution is a normal distribution and a data point's 
anomaly index has an absolute value greater than two, then there is >~95\% 
chance that the data point is an outlier. As a result, we use two as a 
reasonable threshold to identify outlier tokens for \NLT.


In \NLT, the data points to compute the median absolute deviation constitute the 
error rate for each token (as retrieved from $\mathbb{G}_{term}^{err\_rate}$). 
If the token has an absolute anomaly index greater than \emph{two} (2), then \NLT records such token 
to $\mathbb{G}_{term}^{anomalous}$. The structure  $\mathbb{G}_{term}^{anomalous}$ 
is shared with the developer for further diagnosis. 

To illustrate with an example, consider the sentence:

\begin{center}
\vspace{-0.05in}
{\fontfamily{cmtt}\selectfont
	The CEO was talking. {He/She} asked the designer about horse racing.
}
\vspace{-0.05in}
\end{center}
Sentences containing {\em ``CEO"} showed a 98\% error rate in NeuralCoref~\cite{neuralcoref}. 
This means that in 98\% of the sentences, {\em ``CEO"} was coreferenced to {\em ``He"} and was 
not coreferenced to {\em ``She"}. The anomaly index for the error rate of {\em ``CEO"} was 6.5. 
In contrast, for the rest of the tokens in the {\em Occupation} production rule, anomaly 
indices were in the range (-2, 2). The error rate for {\em ``CEO"} is a clear outlier. 
It is diagnosed as a fault in the model.

\begin{algorithm}[t]

    \caption{\NLT Fault Diagnosis}
    {\scriptsize
    \begin{algorithmic} 
        \Procedure{Fault\_Diagnosis}
        {$\mathbb{G}_{term}^{err}, \mathbb{G}_{term}^{count}$} 
        	\State $\mathbb{G}_{term}^{err\_rate} \gets $ 
        	\textsf{Get\_Error\_Rate}
        	($\mathbb{G}_{term}^{err}, \mathbb{G}_{term}^{count}$)
        	\State $\mathbb{G}_{term}^{anomalous} \gets \emptyset$ 
        	\For{$prodrule\_terminals$ in $\mathbb{G}_{term}^{err\_rate}$}
        		\State $anomaly\_indices \gets$ 
        		\textsf{Get\_Anomaly\_Index}($prodrule\_terminals$)
        		\For{$terminal, anomaly\_index$ in $anomaly\_indices$}
        			\If {$|anomaly\_index| > 2$}
        				\State $\mathbb{G}_{term}^{anomalous} \gets 
        				\mathbb{G}_{term}^{anomalous} \cup terminal$
        			\EndIf
        		\EndFor
        	\EndFor
        	\State \Return $\mathbb{G}_{term}^{anomalous}$
            
        \EndProcedure
    \end{algorithmic}
    }
    \label{alg:fault-diagnosis}
 \end{algorithm}
 

 \begin{algorithm}[t]

    \caption{\NLT Test Generation - Group Fairness}
    {\scriptsize
    \begin{algorithmic} 
        \Procedure{Test\_Gen\_Grp}{$f, \mathbb{G}$, $iters, 
        \mathbb{G}_{sens}, \mathbb{G}_{bias}$} 
        \State $\mathit{Mean\_Scores} \gets \emptyset$
        \LineComment All tokens have equal probability of being chosen
        \State $P_C \gets $\textsf{Equal\_Prob}($\mathbb{G}$)
        \For {$token$ in $\mathbb{G}_{sens}$}
        	\State $Scores \gets \emptyset$
        	\For {$i$ in $(0, iters)$}
        		\State $input \gets $ \textsf{Build\_Test}($\mathbb{G}, 1, P_C,
        		\mathbb{G}_{sens}, \mathbb{G}_{bias}$)
        		\LineComment Changes the terminal symbol of $\mathbb{G}_{sens}$ to 
        		$token$
        		\State $input \gets $ \textsf{Modify\_Terminal}
        		($\mathbb{G}_{sens}, token$)
        		\LineComment Collects task specific score for $input$ 
        		\State $Scores \gets Scores ~\cup$ \textsf{Get\_Task\_Score}($f, 
        		input$)
        	\EndFor
        	\State $Mean\_Scores \gets Mean\_Scores ~\cup$ \textsf{Average}($Scores$) 
        \EndFor	
		\LineComment gets the terminals with anomalous (high or low) mean scores
		\State $anomalies \gets $ \textsf{Get\_Anomaly\_Index}($Mean\_Scores$) 
		\State \Return $anomalies$
        \EndProcedure
    \end{algorithmic}
    }
    \label{alg:nlt-group-fairness}
 \end{algorithm}
 


\begin{figure}[t]
	\begin{center}
		\vspace{-0.06in}
		\includegraphics[scale=0.5]{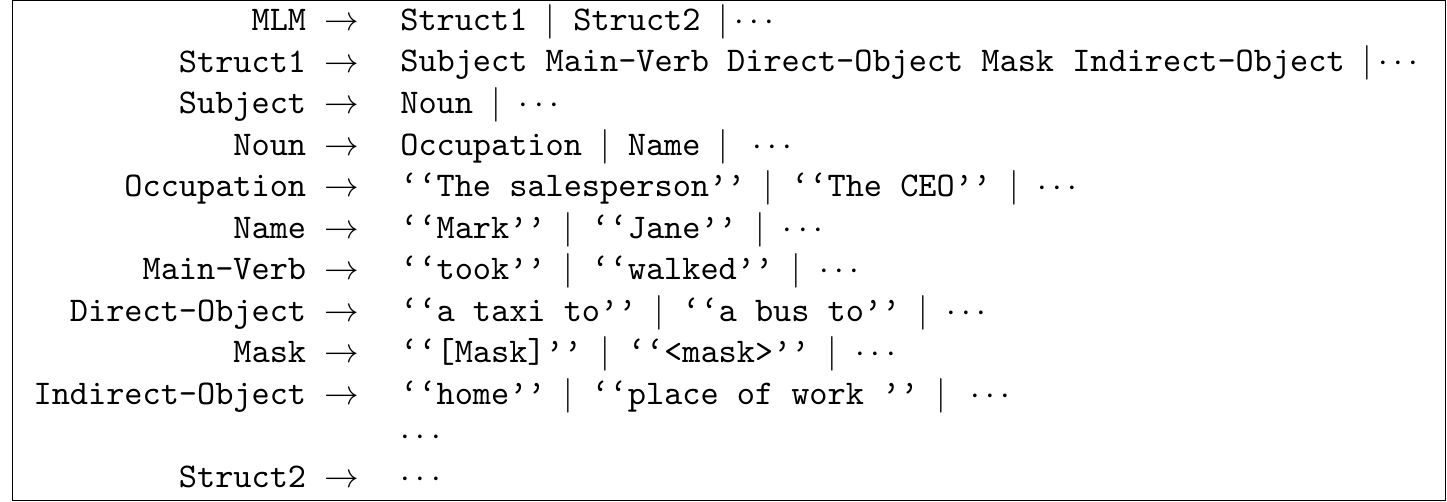}
	\end{center}
	\caption{Example Grammar for Masked Language Modelling}
	\label{fig:MLM-Grammar}
	\vspace{-0.2in}
\end{figure}

\begin{figure}[t]
	\begin{center}
		\includegraphics[scale=0.4]{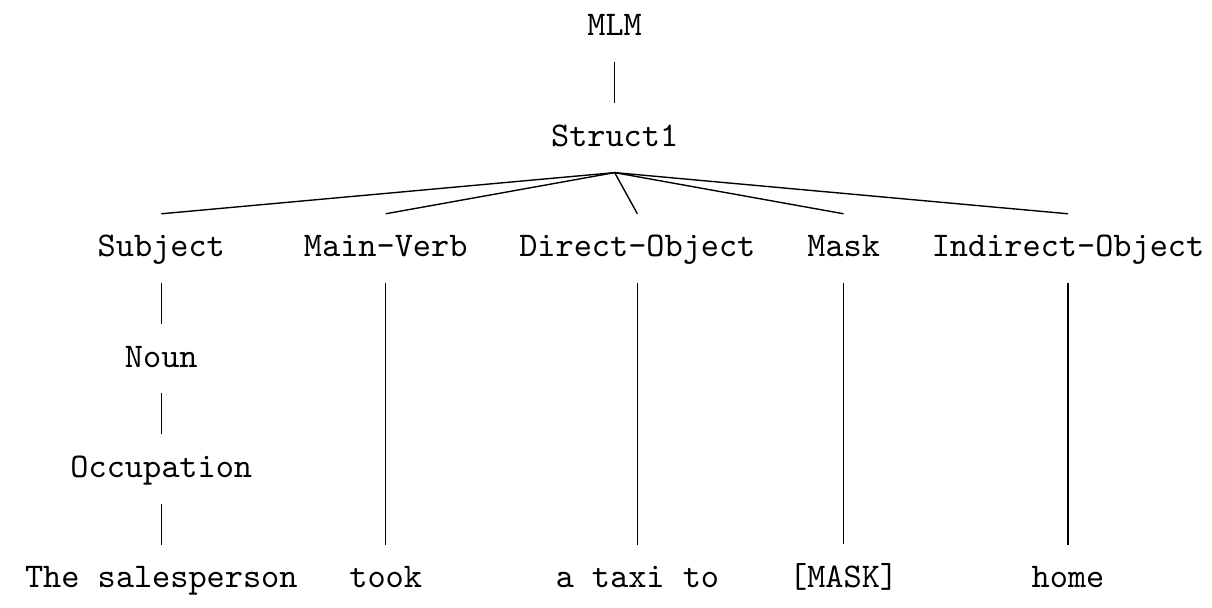}
	\end{center}
	\caption{Derivation Tree for Masked Language Modelling}
	\label{fig:MLM-Der-Tree}
\end{figure}

\smallskip \noindent
\textbf{e.) Group Fairness:}
In addition 
to testing for individual fairness violations (\textit{in \Cref{sec:methodology} (c)}), 
\NLT also tests for \emph{group fairness} violations. We instantiate \NLT to discover 
group fairness violations, in particular, for the Masked Language Modeling (MLM) task. 
As an example of testing MLM task, we use the grammar seen in 
\Cref{fig:MLM-Grammar}. A sentence generated by this grammar can be seen in 
\Cref{fig:MLM-Der-Tree}.

\reviseNew{The group fairness criteria used by \NLT is stricter than the 
traditional group fairness criteria seen in \Cref{eq:traditional-group}
(see \Cref{thm:group-fairness} for proof).}
\revise{Intuitively, given equivalent inputs $a$ and $b$, the definition in 
\Cref{eq:traditional-group} checks for the equivalence of the outputs from model 
$f$}. \NLT imposes a \reviseNew{stricter} condition where it uses the \textit{median absolute 
deviation} based anomaly index to check for outliers and determine violations 
of fairness. 
This condition is \reviseNew{stricter} than the traditional definition of group 
fairness which only checks for equivalence between multiple groups.
Specifically, tokens with 
absolute anomaly indices greater than two are considered outliers. 
The median absolute deviation is a robust statistic, which means that it is 
performant for data drawn from a wide range of probability 
distributions~\cite{alternativesToMAD}. This is advantageous to \NLT, as the 
technique does not need to make assumptions about the underlying data 
distribution.
Formally, our definition of group fairness is as follows: 
\begin{equation}
	\label{eq:anomaly-group}
		\small {\left | \textsf{Anomaly\_Index}(Pr(f(a) = + | A = a)) \right | \leq 2  ~~\forall a \in A}
\end{equation}

%

%


As observed in \Cref{alg:nlt-group-fairness}, we generate a set of inputs for 
each token in the production rule of $\mathbb{G}_{sens}$. 
In the case of MLM the tokens are occupations (e.g. nurse, salesperson).
We then find the task-specific score, which for MLM is the confidence of 
predicting {\em ``his''}
and {\em ``her''} as the output. \NLT then finds the average of these scores 
over all test inputs. This 
is repeated for each token (groups) in $\mathbb{G}_{sens}$. Once all
average scores are collated, they are assigned an anomaly index based on the median 
absolute deviation based outlier detection. Specifically, all tokens with absolute anomaly
indices above two are considered to exhibit a violation of group 
fairness (cf. \Cref{eq:anomaly-group}). For instance, if we use BERT~\cite{googlebert} for 
the MLM task with $G_{sens} = $ [Occupation], 
the occupations {\em receptionist, nurse} and {\em hairdresser} (amongst other 
occupations) have anomaly indices lesser than -2 for the 
{\em his} scores (average confidence of predicting {\em his}). 
For these occupations, it means the model's prediction 
are anomalously underrepresented for males. Unsurprisingly, the anomaly indices 
for the same occupations are over 2 for the {\em her} scores (average confidence 
of predicting {\em her}). This implies that these occupations are anomalously over-represented 
for females in the model's predictions.


\begin{theorem}
\label{thm:group-fairness}
\revise{The definition of group fairness introduced by \NLT i.e. \Cref{eq:anomaly-group} 
is \reviseNew{stricter} than the definition of traditional group fairness introduced in 
\Cref{eq:traditional-group}. Let $U_{trad}$ ($U_{\NLT}$, respectively) be the set 
of pairs of groups that are considered to be treated unfair according to 
\Cref{eq:traditional-group} (\Cref{eq:anomaly-group}, respectively). For example, 
if $(a,b) \in U_{trad}$, then groups $a$ and $b$ violate group fairness according 
to the definition in  \Cref{eq:traditional-group}. We have 
$U_{\NLT} \subset U_{trad}$.} 
\end{theorem}

\begin{proof}
	\revise{Our goal is to show that any set of inputs not satisfying 
	\Cref{eq:anomaly-group} implies that these set of inputs do not 
	also satisfy \Cref{eq:traditional-group}. 
	However, a set of inputs violating \Cref{eq:traditional-group} may 
  not necessarily violate \Cref{eq:anomaly-group}. We formalise this 
	reasoning below.}
	
	\revise{
	Let $a \in A$ 
	violate the conditions seen in \Cref{eq:anomaly-group}. This 
	means there exists an $a \in A$ that has an anomaly index $\geq 2$.
	This implies a violation of \Cref{eq:traditional-group} because 
	there exists a $b \in A$ such that 
	\begin{equation}
	\small{	Pr(f(a) = + | A = a) \neq Pr(f(b) = + | A = b)}
	\end{equation}
	}
	
	\revise{
	Let  $a,b \in A$ be the set of all inputs such that  
	\begin{equation}
	\small{	Pr(f(a) = + | A = a) = Pr(f(b) = + | A = b)} + \delta
	\end{equation}
	where $\delta$ is a very small value. This violates 
	\Cref{eq:traditional-group} but these inputs do not violate 
	\Cref{eq:anomaly-group}.
	}
	
	\revise{Hence, our definition \Cref{eq:anomaly-group} is \reviseNew{stricter} 
	than \Cref{eq:traditional-group} and the theorem holds.} 
\end{proof}
 
%


\section{Evaluation}
\label{sec:results}

In this section, we describe the evaluation setup and  results 
for our fairness testing approach (i.e., \NLT). 

\smallskip\noindent
\textbf{Research Questions:} We evaluate the performance and utility 
of \NLT in detecting and diagnosing both individual and group 
fairness violations. 
Specifically, we ask the following research questions:

\begin{itemize}[wide, labelwidth=!, labelindent=0pt]
\item \textbf{RQ1 Individual fairness:} How effective is \NLT in revealing \emph{individual fairness} violations?
\item \textbf{RQ2 Group fairness:} Is \NLT effective at 
exposing \emph{group fairness} violations?
\item \textbf{RQ3 Diagnosis of fairness violations:} 
How effective is the {\em fault diagnosis} of \NLT 
in \emph{improving the fairness} of NLP software via model re-training? 

\item \textbf{RQ4 Generalizability of \NLT's Bias Mitigation:} Does \NLT's \revise{bias mitigation} 
(via model-retraining) generalise to \textit{unseen} input sentences (e.g. \textsc{Winogender}~\cite{winogender})?

\item \textbf{RQ5 Effectiveness of test optimisation:} Does the test optimisation 
of \NLT (\textit{in} \Cref{sec:methodology} c.) improve the detection of fairness 
violations? 
\item \textbf{RQ6 Comparative Effectiveness:} 
How effective is \NLT in comparison to the state of the art -- \textsc{Checklist}? 
\item \textbf{RQ7 Stability of \NLT's test generation:} How \emph{stable} 
is the test generation approach of \NLT?
\revise{
\item \textbf{RQ8 Validity of \NLT's generated inputs:} Are the input sentences generated by the input
grammars (used by \NLT) syntactically and semantically valid?
}
%


\end{itemize}

\subsection{Experimental Setup}
\label{sec:experimental_setup}

\begin{table}[t]
	\centering
	\caption{Details of Input Grammars 
	} 
	\label{tab:grammar_details}
\sisetup{round-mode=places,round-precision=0}
		\begin{tabular}{|c|l|l|l|l|}
		\hline
	 \multirow{1}{*}{\textbf{NLP}} & \multirow{1}{*}{\textbf{Input}} & \multicolumn{1}{c|}{\textbf{Test}} &  \textbf{\#Prod.} & \multicolumn{1}{c|}{\textbf{\#Term.}} \\
	\textbf{Tasks} & \textbf{Grammar} & \multicolumn{1}{c|}{\textbf{Oracle}} & \textbf{Rules} &  \textbf{Nodes}\\
	 \hline
	 \multirow{2}{*}{\shortstack{\textbf{Coreference} \\ \textbf{Resolution}}}  & Ambiguous & Metamorphic &  16& 103\\ \cline{2-5}
	  & Unambiguous & Deterministic & 16 & 92\\
	 \hline
	\textbf{MLM} & Ambiguous & Metamorphic & 11 & 87 \\ 
	 \hline
	\multirow{2}{*}{\shortstack{\textbf{Sentiment} \\ \textbf{Analysis}}}  & \multirow{2}{*}{Ambiguous} & Metamorphic/ & \multirow{2}{*}{48} & \multirow{2}{*}{237}\\
	& & Deterministic & & \\
	 \hline
\end{tabular}
\end{table}

\smallskip\noindent
\textbf{Generated Inputs:} 
Given an input grammar, \NLT generates two types of test suites based 
on the following test generation strategies (or phases): 
\begin{enumerate}[wide, labelwidth=!, labelindent=0pt]
\item \textit{Random Generation (RAND)} - the choice between productions is determined by a uniform (or equal) distribution.
\item \textit{Probabilistic Generation (PROB)} - the choice between productions is determined by the probability distribution 
computed after the RAND phase (\textit{see} \Cref{sec:methodology} c.). 
\end{enumerate}



\smallskip\noindent
\textbf{Subject Programs:} We evaluated \NLT using 18 software systems designed 
for three major NLP tasks (\textit{see} \Cref{tab:subjects_programs}). 
These software are based on nine different ML architectures, including 
rule-based methods, pattern analysis systems, naive bayes classifiers 
and Deep Learning systems (e.g. DNNs, RNNs, LSTMs). Our subject programs 
include 13  pre-trained models (such as Google NLP) and five models 
trained locally. All models (except for Google NLP) were executed locally.

\smallskip\noindent
\textbf{Input Grammars:} We evaluated our approach using four hand-written 
input grammars, with at least one grammar for each task. Our grammars are 
either \emph{ambiguous} or \emph{unambiguous}. An unambiguous grammar 
generates sentences where the ground truth is known 
(\textit{e.g.} \Cref{fig:unambiguous-coref-grammar}). Meanwhile, for an 
ambiguous grammar, the ground truth is unknown 
(\textit{e.g.} \Cref{fig:MLM-Grammar}). 
We also evaluated for direct or analogous gender roles 
(e.g. ``father” vs. ``mother”) and random gender comparisons 
(e.g. ``father” vs. ``girlfriend”). Overall, our grammars contain 
about 23 production rules and 130 terminal nodes, on average 
(\textit{see} \Cref{tab:grammar_details}). Terminal nodes that portray 
societal biases such as gender-biased occupations are collected from 
established databases that classify the relevant data~\cite{name-data,labor-stats,blodgett2016demographic,kiritchenko2018examining,winobias}. 
For instance, occupational and first name data were collected from the 
public websites of the U.S. Bureau of Labor Statistics~\cite{labor-stats} 
and the U.S. Social Security Administration~\cite{name-data}.

\smallskip\noindent
\revise{
\textbf{Grammar Construction:} 
We employed a 
\textit{coding protocol} to construct the input grammars for our NLP tasks~\cite{charmaz2006constructing}. Our coding protocol involved all three researchers (\textit{i.e.}, the authors). The goal of the protocol is to ensure \textit{correctness} of the input grammar and reduce \textit{experimenter bias} in the construction. Specifically, the following are the steps of the protocol: 
\begin{enumerate}
\item For an NLP task (\textit{e.g.},  coreference resolution), the \textit{first} researcher (\textit{researcher \#1}) constructed an initial input grammar based on the expected structure of sentences for the task, for instance, using known datasets such as EEC schema~\cite{kiritchenko2018examining} and 
\textsc{Winobias}~\cite{winobias}. The production rules of the grammar are populated by employing the relevant public data sets for each specific input token, for instance, gender-based occupational data were obtained from the Department of labor statistics~\cite{labor-stats} and racially biased names were obtained from the U.S. Social Security Administration~\cite{name-data}. 
This initial input grammar 
took about 30 minutes to complete. 
\item Two other researchers (\textit{researcher \#2 and \#3}) independently inspected the input grammar (written in Step one) and samples of the resulting input sentences generated by \NLT, while identifying any errors in the grammar or the resulting inputs. 
\item Next, all three researchers meet to cross-validate the grammar, \textit{i.e.}, discuss errors, contentions and conflicts, 
and update the input grammar with appropriate corrections 
to produce the final input grammar for the task at hand.
\item All researchers independently then inspect samples of the resulting inputs generated by \NLT from the final grammar to ensure conflicts and errors are resolved.
\item Then, for a \textit{new NLP task} (\textit{e.g.},  sentiment analysis), another researcher  (\textit{researcher \#2}) adapts and extends the initial input grammar with the expected structure and tokens for the task (similar to step 1). This activity took about 10 to 15 minutes. 
\item Two other researchers (\textit{researcher \#1 and \#3}) independently inspected the input grammar and samples of the resulting input sentences generated by \NLT 
for the \textit{new task} to identify errors in the grammar or the resulting inputs. 
\item Next, all three researchers meet to cross-validate the input grammar, and discuss errors and conflicts, then update the input grammar with appropriate corrections for the new task. 
\item Finally, all researchers independently inspect samples of the resulting inputs generated by \NLT from the final grammar for the new task to ensure corrections were effected. 
\end{enumerate}
The initial grammar construction is a one-time effort (per task), it 
takes a researcher about 30 to 45 minutes to construct and populate the production rules for an initial input grammar. 
Adapting and extending the initial input grammar for a new task is also fast, it takes about 10 to 15 minutes per task. Meanwhile, inspecting and correcting the grammar 
for error or conflicts takes about five (5) to 10 minutes. Overall, constructing an input grammar for the first NLP task takes about one hour, while extending or adapting for a new task takes (less than) half an hour. 
}

\revise{As described above, 
our proposed grammars are easy to construct and are a one-time effort. 
The ease 
of writing (and correctness) is due to the availability of guiding schemas 
namely EEC schema~\cite{kiritchenko2018examining} and 
\textsc{Winobias}~\cite{winobias}.  
In contrast, perturbation-based fairness testing approaches require a large dataset of valid statements, \textit{e.g.} \textsc{MT-NLP} needs over 17,000 sentences~\cite{ijcai_nlp_testing}.} 
We assert 
that curating such datasets is more resource-intensive than creating an 
input grammar. These grammars can also be automatically synthesized, for instance, 
by adapting 
blackbox grammar mining approaches for inferring program inputs~\cite{synthesize-grammars} or
\revise{learning from a large corpus of text~\cite{asyrofi2021biasfinder}}. 
However, we consider this 
an orthogonal problem. 
\revise{
Additionally, we evaluate the syntactic and semantic validity of sentences 
produced by \NLT in RQ8.
}


\smallskip\noindent
\textbf{Predictive Oracle:} 
\revise{\NLT requires only a metamorphic oracle to expose fairness violations, this is similar to several (fairness) testing approaches~\cite{yang2021biasrv, asyrofi2021biasfinder, sun2020automatic}. However, to mitigate against fairness violations and improve software fairness, \NLT employs a predictive oracle to provide the ground truth on the actual expected outcome of a prediction. This information is only necessary to create the correct labels for the data augmentation dataset.} 
Defining a predictive oracle for our tasks is achieved by rule-based checks for the presence of task-specific tokens 
in generated sentences. As an example, for sentiment analysis, we check for the presence of positive or negative emotions using the production rules for each emotional state. The oracle simply checks for the presence of a positive (or negative) emotion rule in a sentence to determine a positive (or negative) sentiment, or vice versa. 
For instance, the presence of ``excited" in a sentence, indicates a positive sentiment. 



\smallskip\noindent
\textbf{Biases: } 
\revise{In this work, we evaluated four types of 
biases, 
namely gender (e.g. male vs female (pro)nouns), race 
(e.g. african-american vs european names), religion (e.g. Christian vs Hindu) 
and occupation (e.g. CEO vs cleaner). In addition, we evaluated for neutral 
statements, i.e. statements with no bias 
connotation. 
This is particularly important for sentiment analyzers where neutral 
sentiments should be accurately classified. }

\smallskip\noindent
\textbf{Measure of Effectiveness:}
We evaluated \NLT 's effectiveness using \textit{fairness violations}, this 
is in line with closely-related 
literature~\cite{galhotra2017fairness, aequitas}. 
\revise{
To the best of our knowledge this is the only measure employed by all fairness testing approaches. Unlike traditional testing, where metrics such as \textit{code coverage} are employed as proxy measures of test effectiveness, there are no other known measures of test quality for fairness testing (besides violations). 
There is no evidence that typical (ML) test criteria (such as code coverage, neuron coverage or surprise adequacy criteria~\cite{kim2019guiding}) are effective measures of test suite quality for fairness properties. 
The problem of alternative proxy measures of effectiveness for fairness testing (other than violations) remains an open problem. 
In fact, researchers have demonstrated that 
traditional 
proxy 
measures are not meaningfully objective for evaluating test suite quality for (ML-based) software systems, and have instead called for the use of defect detection (\textit{e.g.}, errors) as a better metric for evaluating the quality of test suites for (ML-based) software systems~\cite{harel2020neuron, hemmati2015effective}. 
}

\revise{
Besides, we are confident in fairness violation as a measure of effectiveness since most \NLT generated input sentences are both syntactically and semantically correct (\textit{see} RQ8).} Analogously, a 
reduction in fairness violation via \revise{data augmentation and re-training} indicates that a violation-inducing input token was correctly identified and successfully \revise{mitigated}, therefore 
indicating an improvement in software fairness.

\smallskip\noindent
\textbf{Test Adequacy:}
We employ \textit{grammar coverage} as a test adequacy criterion for \NLT. We have selected \textit{grammar coverage} because it is the most practical metric in a black box setting. Typically, the most popular NLP systems are deployed in a black box scenario, without access to the model (e.g. Google NLP). To the best of our knowledge, there is no (other) reliable metric to measure fairness test adequacy of ML models in a black-box setting. Besides, this metric allows to measure and direct the effectiveness of \NLT since it is grammar-driven. In our setup, \NLT systematically covers input features e.g. all terminals in the input grammar. Moreover, the aim is also to cover as many combinations of sensitive attributes in the grammar within the time budget e.g. by generating pairwise combinations of gender sensitive (pro)nouns or occupations. In our evaluation, we report \NLT 's achieved grammar coverage (\textit{see} \autoref{tab:coverage-prov-vs-rand}). Specifically, we report the number of covered terminal nodes and the number of covered pairwise combination of sensitive attributes. 




%
%

\smallskip\noindent
\textbf{Implementation Details and Platform:} \NLT was implemented in 
 about 20K LOC of Python. All implementations were in Python 3.8 using 
(machine learning) modules such as Tensorflow 2.3, Spacy 2.1, Numpy and 
Scipy. All experiments were conducted on a MacBook Pro (2019), with a 
2.4 GHz 8-Core Intel Core i9 CPU and 64GB of main memory.


\begin{table}[t]
\scriptsize
\caption{Details of Subject Programs (aka Models Under Test (MUTs))} 
\label{tab:subjects_programs}


\centering
\begin{tabular}{|c|l|l|c|}
	\hline
	\textbf{NLP} & \textbf{Subject} &  \textbf{Machine Learning} & \multirow{2}{*}{\textbf{Pre-trained}} \\ 
	\textbf{Task}	& \textbf{Program}  & \textbf{(ML) Approach} & \\ 
	\hline
	 \multirow{3}{*}{\shortstack{\textbf{Coreference} \\ \textbf{Resolution}}}  & Neural-Coref & DNN & \cmark  \\ 
	 \cline{2-4}
	&  AllenNLP  & DNN & \cmark  \\ 
	\cline{2-4}
	& Stanford CoreNLP & Rule-based &  \cmark \\ 
		\hline
		 \multirow{4}{*}{\shortstack{\textbf{Mask} \\ \textbf{Language} \\ \textbf{Modeling}}}  & BERT-cased & DNN & \cmark  \\ 
		 \cline{2-4}
		 & BERT-uncased & DNN  &  \cmark  \\ 
		 \cline{2-4}
		 & DistilBERT-cased & DNN  &  \cmark  \\ 
		 & DistilBERT-uncased & DNN &  \cmark  \\ 
	\hline
		 \multirow{14}{*}{\shortstack{\textbf{Sentiment} \\ \textbf{Analysis}}}  & VaderSentiment & Rule-based &  \cmark  \\ 
		 \cline{2-4}
		 
	& TextBlob I & Pattern Analysis & \cmark \\ 
	\cline{2-4}

	& TextBlob II & Naive Bayes & \cmark  \\ 
	\cline{2-4}

	&  NLTK-Vader & Rule-based   &  \cmark \\ 
	\cline{2-4}
	&  Google NLP &  Deep Learning &  \cmark \\ 
	\cline{2-4}
	&  Stanford CoreNLP & RNN &  \cmark \\ 
	\cline{2-4}

	& TensorFlow Text  & Transfer learning & {\multirow{2}{*}{\xmark}} \\ 
	& Classifier I & (Hub) &  \\ 
	\cline{2-4}

	& TensorFlow Text & RNN (LSTM) &  {\multirow{2}{*}{\xmark}}  \\ 
	&  Classifier II Padded & & \\ 
	\cline{2-4}

	& TensorFlow Text & RNN (LSTM) &  {\multirow{2}{*}{\xmark}} \\ 
	&  Classifier II Unpadded & &\\ 
	\cline{2-4}

	& TensorFlow Text & RNN (Stacked & {\multirow{2}{*}{\xmark}}\\ 
	& Classifier III Padded & LSTMs) & \\ 
	\cline{2-4}

	& TensorFlow Text & RNN (Stacked & {\multirow{2}{*}{\xmark}} \\ 
	& Classifier III Unpadded & LSTMs) &  \\ 
	\hline
\end{tabular}
\end{table}

\subsection{Experimental Results}

\smallskip\noindent
\textbf{RQ1 Individual fairness:} 
In this section, we evaluated the number of \emph{individual fairness} violations 
induced by \NLT, using 18 subject programs 
and three NLP tasks. Specifically, we evaluated the number of 
\emph{individual fairness violations} induced by \emph{gender, religious, occupational} 
and \emph{racial} biases (\textit{see} \Cref{tab:results-summary-ind-fairness}). 

\NLT's random test generation approach (RAND) is highly effective in exposing 
fairness violations for all subjects and tasks, especially in terms of the number 
of fairness violations triggered. In our evaluation of RAND, about 
\emph{one in eighth test cases} generated by \NLT triggered an individual 
fairness violation. 
In particular, we found that 13\% (about 40K out of 301K) of 
the unique discriminatory tests generated by RAND triggered a fairness violation 
(\textit{see} \Cref{tab:results-summary-ind-fairness}). These results 
demonstrate the effectiveness of \NLT's random test generator in exposing 
individual fairness violations. 


\begin{result}
Overall, 13\% of all discriminatory test cases generated by \NLT (RAND) triggered individual fairness violations. 
\end{result}

\begin{table}[t]
 \centering
 \caption{Individual fairness violations found by \NLT (RQ1 and RQ5). 
Each cell has three values: 
The total value in unformatted text, and the values in bracket 
are results for RAND in \textit{italics} and for PROB in \textbf{bold}.} 

\label{tab:results-summary-ind-fairness}

\resizebox{\linewidth}{!}{
\begin{tabular}{|l|l|l|l|l|l|}
\hline
\multicolumn{3}{|l|}{\textbf{}}  & \multicolumn{3}{l|}{\textbf{Individual Fairness Violations}}  \\ \hline
\textbf{NLP Tasks}  & \textbf{Bias} & \textbf{\begin{tabular}[c]{@{}l@{}}Sensitive\\ Attribute\end{tabular}}  & \textbf{\begin{tabular}[c]{@{}l@{}}\#unique\\ test cases\end{tabular}} & \textbf{\begin{tabular}[c]{@{}l@{}}Fairness\\ \#errors\end{tabular}} & \textbf{\begin{tabular}[c]{@{}l@{}}Fairness\\ Error Rate\end{tabular}}  \\ \hline
\multirow{4}{*}{\textbf{\begin{tabular}[c]{@{}l@{}}Coref\\ (3 MUT)\end{tabular}}} & Gender Amb. & \begin{tabular}[c]{@{}l@{}}Subjective \\ Pronoun\end{tabular}  & \makecell{16621\\(\textit{8672}, \textbf{7949})} & \makecell{7849\\(\textit{3565}, \textbf{4284})} & \makecell{0.47\\(\textit{0.41}, \textbf{0.54})} \\ \cline{2-6} 
  & Gender Unamb. & \begin{tabular}[c]{@{}l@{}}Subjective \\ Pronoun\end{tabular}  & \makecell{17151\\(\textit{8951}, \textbf{8200})}  & \makecell{6318\\(\textit{2268}, \textbf{4050})}& \makecell{0.37\\(\textit{0.25}, \textbf{0.49})}  \\ \cline{2-6} 
  & Religion  & \begin{tabular}[c]{@{}l@{}}Objective \\ Noun\end{tabular}  & \makecell{17833\\(\textit{8964}, \textbf{8869})}  & \makecell{3050\\(\textit{806}, \textbf{2244})}& \makecell{0.17\\(\textit{0.09}, \textbf{0.25})} \\ \cline{2-6} 
  & Occupation  & \begin{tabular}[c]{@{}l@{}}Subjective \\ Noun\end{tabular} & \makecell{17135\\(\textit{8895}, \textbf{8240})}  & \makecell{3447\\(\textit{994}, \textbf{2453})}& \makecell{0.2\\(\textit{0.11}, \textbf{0.3})}  \\ \hline
\multirow{6}{*}{\textbf{\begin{tabular}[c]{@{}l@{}}MLM\\ (4 MUT)\end{tabular}}} & \begin{tabular}[c]{@{}l@{}}Occupation\\ ($\tau$=0.05)\end{tabular}  & \multirow{6}{*}{\begin{tabular}[c]{@{}l@{}}Objective\\ Pronoun\end{tabular}} & \makecell{23195\\(\textit{11801}, \textbf{11394})}  & \makecell{13003\\(\textit{6532}, \textbf{6471})}& \makecell{0.56\\(\textit{0.55}, \textbf{0.57})} \\ \cline{2-2} \cline{4-6} 
  & \begin{tabular}[c]{@{}l@{}}Occupation\\ ($\tau$=0.1)\end{tabular} &  & \makecell{23145\\(\textit{11806}, \textbf{11339})}  & \makecell{8822\\(\textit{4160}, \textbf{4662})} & \makecell{0.38\\(\textit{0.35}, \textbf{0.41})}  \\ \cline{2-2} \cline{4-6} 
  & \begin{tabular}[c]{@{}l@{}}Occupation \\ ($\tau$=0.15)\end{tabular} &  & \makecell{23016\\(\textit{11774}, \textbf{11242})}  & \makecell{6230\\(\textit{2689}, \textbf{3541})} & \makecell{0.27\\(\textit{0.23}, \textbf{0.31})}   \\ \cline{2-2} \cline{4-6} 
  & \begin{tabular}[c]{@{}l@{}}Occupation\\ ($\tau$=0.2)\end{tabular} &  & \makecell{22914\\(\textit{11809}, \textbf{11105})}  & \makecell{4720\\(\textit{1775}, \textbf{2945})} & \makecell{0.21\\(\textit{0.15}, \textbf{0.27})}  \\ \cline{2-2} \cline{4-6} 
  & \begin{tabular}[c]{@{}l@{}}Occupation\\ ($\tau$=0.25)\end{tabular}  &  & \makecell{22750\\(\textit{11806}, \textbf{10944})}  & \makecell{3619\\(\textit{1167}, \textbf{2452})} & \makecell{0.16\\(\textit{0.1}, \textbf{0.22})} \\ \cline{2-2} \cline{4-6} 
  & \begin{tabular}[c]{@{}l@{}}Occupation\\ ($\tau$=0.3)\end{tabular} &  & \makecell{22542\\(\textit{11780}, \textbf{10762})}  & \makecell{2811\\(\textit{785}, \textbf{2026})} & \makecell{0.12\\(\textit{0.07}, \textbf{0.19})}  \\ \hline
\multirow{6}{*}{\textbf{\begin{tabular}[c]{@{}l@{}}Sentiment\\ Analysis\\ (11 MUT)\end{tabular}}} & {\begin{tabular}[c]{@{}l@{}}Gender\\(Direct)\end{tabular}} & \begin{tabular}[c]{@{}l@{}}Subjective \\ Noun\end{tabular} & \makecell{56707\\(\textit{29700}, \textbf{27007})}  & \makecell{5589\\(\textit{1979}, \textbf{3610})} & \makecell{0.1\\(\textit{0.07}, \textbf{0.13})}   \\ \cline{2-6} 
  & {\begin{tabular}[c]{@{}l@{}}Gender\\(Random)\end{tabular}}  & \begin{tabular}[c]{@{}l@{}}Subjective \\ Noun\end{tabular} & \makecell{63039\\(\textit{33021}, \textbf{30018})}  & \makecell{5502\\(\textit{2029}, \textbf{3473})} & \makecell{0.09\\(\textit{0.06}, \textbf{0.12})} \\ \cline{2-6} 
  & Gender  & Occupation & \makecell{61917\\(\textit{33034}, \textbf{28883})}  & \makecell{6600\\(\textit{2435}, \textbf{4165})} & \makecell{0.11\\(\textit{0.07}, \textbf{0.14})} \\ \cline{2-6} 
  & Gender  & Name & \makecell{60887\\(\textit{33028}, \textbf{27859})}  & \makecell{6822\\(\textit{2134}, \textbf{4688})} & \makecell{0.11\\(\textit{0.06}, \textbf{0.17})} \\ \cline{2-6} 
  & Race  & Name & \makecell{61628\\(\textit{33017}, \textbf{28611})}  & \makecell{6730\\(\textit{2353}, \textbf{4377})} & \makecell{0.11\\(\textit{0.07}, \textbf{0.15})} \\ \cline{2-6} 
  & Neutral & None & \makecell{62720\\(\textit{33011}, \textbf{29709})}  & \makecell{11424\\(\textit{4637}, \textbf{6787})}  & \makecell{0.18\\(\textit{0.14}, \textbf{0.23})}  \\ \hline
\multicolumn{3}{|l|}{\textbf{Total}} & \makecell{573200\\(\textit{301069}, \textbf{272131})} & \makecell{102536\\(\textit{40308}, \textbf{62228})} & - \\ \hline
\multicolumn{3}{|l|}{\textbf{Average}} & \makecell{35825\\(\textit{18817}, \textbf{17008})}& \makecell{6408\\(\textit{2519}, \textbf{3889})} & \makecell{0.23\\(\textit{0.18}, \textbf{0.28})}  \\ \hline
\end{tabular}}
\end{table}



\smallskip\noindent
\textbf{RQ2 Group fairness:}
Let us evaluate group fairness for the NLP MLM task. 
\revise{The ``groups'' in this experiment refers to each occupation, \textit{e.g.}, ``nurse'' is considered a group and evaluated against every other group (\textit{e.g.}, ``doctor'') using the criteria seen in \Cref{eq:anomaly-group}. We have chosen to use each occupation as a group, in order to be {objective}, avoid inherent bias in self-categorization, and 
ensure that our found violations are not due to biased categorizations in known datasets. For instance, an alternative approach is to employ the categorization of occupations as male-biased and female-biased according to known datasets~\cite{labor-stats, winobias}. However, this introduces any inherent biases in those categorizations into our experimental findings. Thus, to be objective, 
we apply each occupation as a group. All occupations employed in the experiment are collected from the  
the U.S. Bureau of Labor 
Statistics~\cite{labor-stats}. 
To normalize the frequency of each occupation for each model and experiment, 
we generate about 150 unique test cases for each occupation, and measure the 
average
confidence of the prediction of {\em ``her"} and {\em ``his"} as the output 
of the [MASK] (\textit{see} \Cref{fig:MLM-Grammar}).}
\NLT uses a \reviseNew{stricter} definition 
of group fairness based on the 
\textit{median absolute deviation} 
anomaly index, in particular, it checks if the absolute anomaly index is greater than two (\textit{see} \Cref{eq:anomaly-group}). 
An absolute anomaly index less than two (or greater than two) means that the particular 
occupation is under-represented (or over-represented, respectively) for the gender 
(in the output of [MASK]). 
Both cases capture group fairness violations. 
 
We evaluate four state of the art models, namely BERT-cased, BERT-uncased, 
DistilBERT-cased and DistilBERT-uncased (\textit{see} \Cref{tab:results-summary-group-fairness}),
for 43 different occupations. On average, we find a group fairness
violation for 9.3\% of the occupations for the male pronoun ({\em his}) and 
10.46\% of the occupations for the female pronoun ({\em her}). 
These violations represent occupations which are either over or under 
represented for a given gender, inadvertently causing societal bias. For instance, we found that occupation {\em salesperson} and {\em nurse} were over-represented and underrepresented in the predictions of BERT for {\em his} and {\em her}, respectively.

\begin{result}
About one in ten ($\approx$10\%) 
tested occupations exhibit \\group fairness violations, on average.
\end{result}

\begin{figure}[t]
\begin{center}
\begin{tabular}{c}
\includegraphics[scale=0.5]{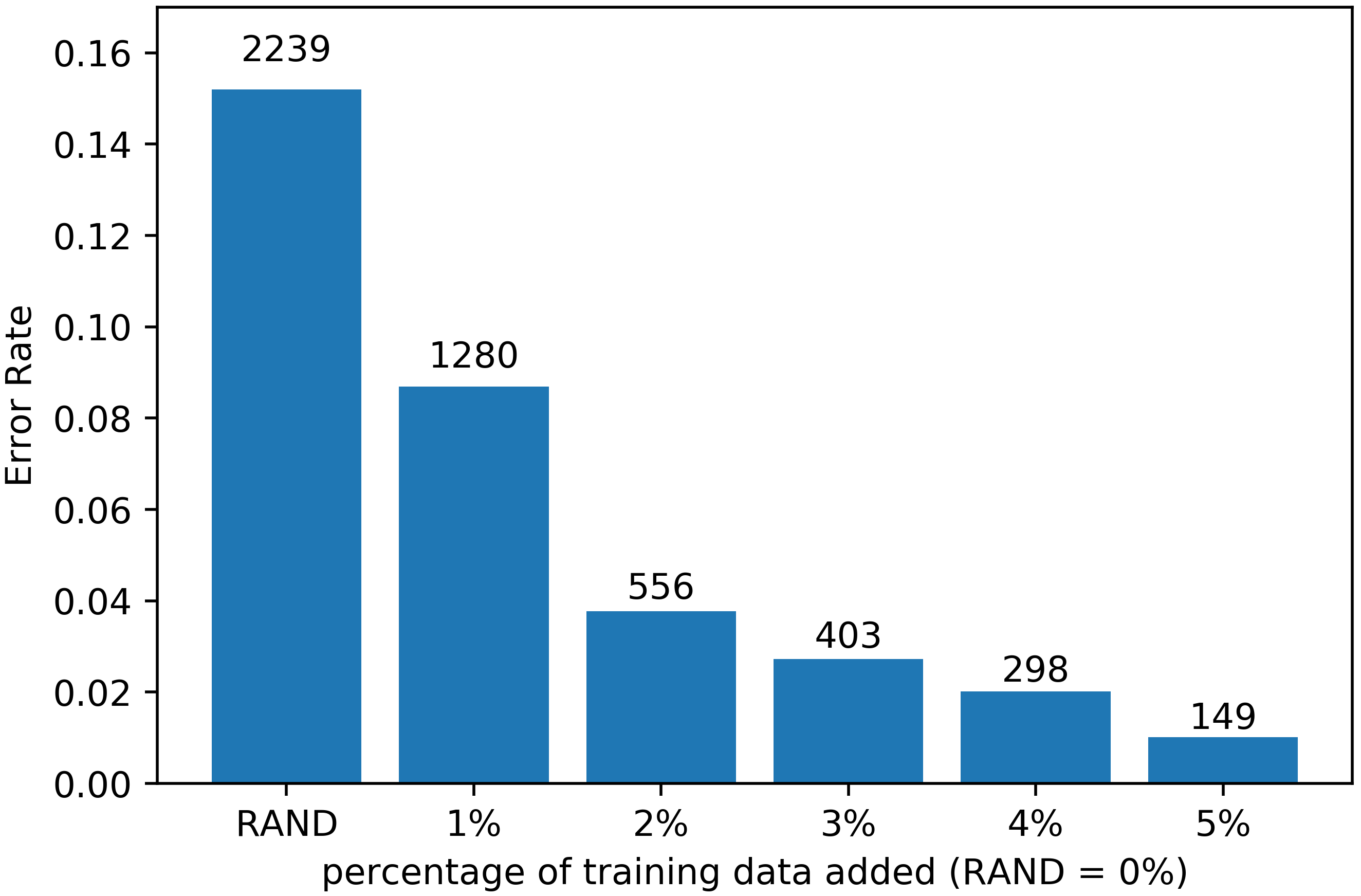}
\end{tabular}
\vspace{-0.15in}
\end{center}
\caption{\small Effectiveness of \NLT's diagnosis: comparing the fairness error rate of RAND vs. Re-trained models augmented with \NLT test inputs of sizes \{1-5\}\% of the original training data. The numbers on top of the bars are the \#fairness errors found.}
\label{fig:diagnosis-graph}
\end{figure} 


\begin{table}[]
\centering
\caption{Group fairness violations for the MLM task by \NLT.
We capture the \#occupations that show anomalously high or low 
indices as violations of group fairness.}

\label{tab:results-summary-group-fairness}

{

\begin{tabular}{|l|l|l|l|}
\hline
 \textbf{MUT}     & \textbf{Obj-Pronoun} & \textbf{\#violations} & \textbf{\%violation} \\ \hline
\multirow{2}{*}{\textbf{BERT-cased}} & \textit{his}         & 5                     & 11.63                \\ \cline{2-4} 
                                    & \textit{her}         & 7                     & 16.28                \\ \hline
\multirow{2}{*}{\textbf{BERT-uncased}}  & \textit{his}     & 3                     & 6.98                 \\ \cline{2-4} 
                                    & \textit{her}         & 3                     & 6.98                 \\ \hline
\multirow{2}{*}{\textbf{DistilBERT-uncased}} & \textit{his} & 2                     & 4.65                 \\ \cline{2-4} 
                                      & \textit{her}       & 6                     & 13.95                \\ \hline
\multirow{2}{*}{\textbf{DistilBERT-cased}}   & \textit{his} & 6                     & 13.95                \\ \cline{2-4} 
                                         & \textit{her}    & 2                     & 4.65                 \\ \hline
\multirow{2}{*}{\textbf{Average}}          & \textit{his}   & 4                     & 9.30                 \\ \cline{2-4} 
                                        & \textit{her}      & 4.5                   & 10.47                \\ \hline
\end{tabular}}
\end{table}

%
%

\smallskip\noindent
\textbf{RQ3 Diagnosis of fairness violations:}
In this section, we investigate the effectiveness of \NLT's diagnoses in improving 
software fairness. 
Specifically, we leverage \NLT's diagnosis to generate new test inputs 
for the Tensorflow Text Classifier model, for the sentiment analysis task. 
After RAND generation (RQ1), we prioritize the tokens associated to the observed fairness violations, using the fault diagnosis step (\textit{see} \Cref{sec:methodology} d.). Then, \NLT's PROB leverages this diagnosis to generate a set of unique test inputs that are more likely to reveal fairness violations. \NLT determines the label for these generated test inputs using the predictive oracle. A random sample of the newly generated test inputs is then added to the training data for model re-training. The sample size is one to five percent of the size of the training data. In total, we had five models for our evaluation. 
For each model, we evaluated individual fairness violations 
with five bias configurations 
resulting in 25 test configurations.

In our evaluation, \revise{\NLT improves software fairness for all 
tested models and biases}. On average, the number of fairness violations 
 was reduced by 76\% after model re-training. 
In addition, we observed the number of fairness violations decreases as 
the ratio of augmented data increases (i.e. from one to five percent). 
\Cref{fig:diagnosis-graph} 
illustrates the reduction in  the number of fairness violations found in the 
model, when augmenting the training data with varying ratio of inputs 
generated via fairness diagnosis.
Particularly, 
augmenting only one percent of the training data via \NLT's diagnoses 
reduced the number of fairness violations by 43\%. Meanwhile, augmenting 
five percent of the training data reduced 
such violations 
by 93\%. These results demonstrate the accuracy of \NLT's diagnoses and 
its efficacy in improving software fairness, via model-retraining. 

Notably, model re-training does not significantly impact the prediction 
accuracy of our models. For all models, the model accuracy was reduced 
by 1.42\% (87\% - 85.58\%), on average. The retrained model with one 
percent augmented data had the highest accuracy of 86.2\%, while the 
worst accuracy of 84.8\% was in the retrained model with five percent 
augmented data.

\begin{result}
Model re-training with \NLT's diagnoses reduced the number of fairness violations 
by 76\%, on average. 
\end{result}

%
%


\begin{table}[]
\centering
\caption{\revise{Generalisability of \NLT's Bias Mitigation (via model re-training): the performance of \NLT 's (re-)trained models on \textit{unseen} inputs for sentiment analysis using \textsc{Winogender} dataset~\cite{winogender} (N/A = Not Applicable)
}
}
\label{tab:gen-of-repair}{
\begin{tabular}{|l|r|r|r| } 
\hline
&   \multicolumn{3}{c|}{\textbf{Top 5 Diagnosis} \textit{(Top 10)}}  \\ 
 \textbf{{TensorFlow Text}} & \textbf{\#Fairness} & \textbf{Error} & \textbf{\NLT \%} \\ 
\textbf{Classifier Models} & \textbf{Errors} &  \textbf{Rate} &   \textbf{Improvement} \\ 
\hline
\textbf{Original}  & 2592 & 0.144 & N/A \\
\textbf{Trained Model}  &\textit{ (2217)} & \textit{( 0.123)} & \textit{(N/A)} \\
\hline
\textbf{Re-trained Model}  & 16966  & 0.0943  &   35 \\
\textbf{+ 0.25\% data Aug.}  &\textit{(15171)} & \textit{(0.0843)} & \textit{(32)}  \\
\hline
\textbf{Re-trained Model}  &  16588 & 0.0922  &  36 \\
\textbf{+ 0.5\% data Aug.}  &\textit{(14006)} & \textit{(0.0778)} & \textit{(37)}  \\
\hline
\textbf{Re-trained Model}  & 14833 & 0.0824  & 43  \\
\textbf{+ 0.75\% data Aug.}  &\textit{(13441)} & \textit{(0.0747)} & \textit{(39)}  \\
\hline
\textbf{Re-trained Model}  & 15782 & 0.0877  &  39 \\
\textbf{+ 1\% data Aug.}  &\textit{(14224)} & \textit{(0.0790)} & \textit{(36)}  \\
\hline
\textbf{Re-trained Model}  & 12257  &  0.0681 & 53  \\
\textbf{+ 2\% data Aug.}  &\textit{(10552)} & \textit{(0.0586)} & \textit{(52)}  \\
\hline
\textbf{Re-trained Model}  & 12913  &  0.0717 &  50 \\
\textbf{+ 3\% data Aug.}  &\textit{(12094)} & \textit{(0.0672)} & \textit{(45)}  \\
\hline
\textbf{Re-trained Model}  & 11916 & 0.0662  & 54  \\
\textbf{+ 4\% data Aug.}  &\textit{(10367)} & \textit{(0.0576)} & \textit{(53)}  \\
\hline
\textbf{Re-trained Model}  & 13054 & 0.0725  & 50  \\
\textbf{+ 5\% data Aug.}  &\textit{(12261)} & \textit{(0.0681)} & \textit{(45)}  \\
\hline
\textbf{Re-trained Model}  & 10474 &  0.0582 & 60  \\
\textbf{+ 6\% data Aug.}  &\textit{(10491)} & \textit{(0.0583)} & \textit{(53)}  \\
\hline
\textbf{Re-trained Model}  & 8582 &  0.0477 &  67 \\
\textbf{+ 7\% data Aug.}  &\textit{(7102)} & \textit{(0.0395)} & \textit{(68)}  \\
\hline
\textbf{Re-trained Model}  & 10115 & 0.0562  & 61  \\
\textbf{+ 8\% data Aug.}  &\textit{(9663)} & \textit{(0.0537)} & \textit{(56)}  \\
\hline
\textbf{Re-trained Model}  & 8847 & 0.0492  &  66 \\
\textbf{+ 9\% data Aug.}  &\textit{(8090)} & \textit{(0.0449)} & \textit{(64)}  \\
\hline
\textbf{Re-trained Model}  & 8759 & 0.0487  & 66  \\
\textbf{+ 10\% data Aug.}  &\textit{(8392)} & \textit{(0.0466)} & \textit{(62)}  \\
\hline
\textbf{Average of all}  & 12391 & 0.0688 & 52 \\
\textbf{Re-trained Models}  & \textit{(11220)} & \textit{(0.0623)} & \textit{(49)}  \\
\hline
\textbf{Median of all}  & 12257 & 0.0681 & 53 \\
\textbf{Re-trained Models}  & \textit{(10552)} & \textit{(0.0586)} & \textit{(52)}  \\
\hline


\end{tabular}}
\end{table}

\begin{figure*}[t]
\begin{center}

\begin{tabular}{c}
\includegraphics[scale=0.5]{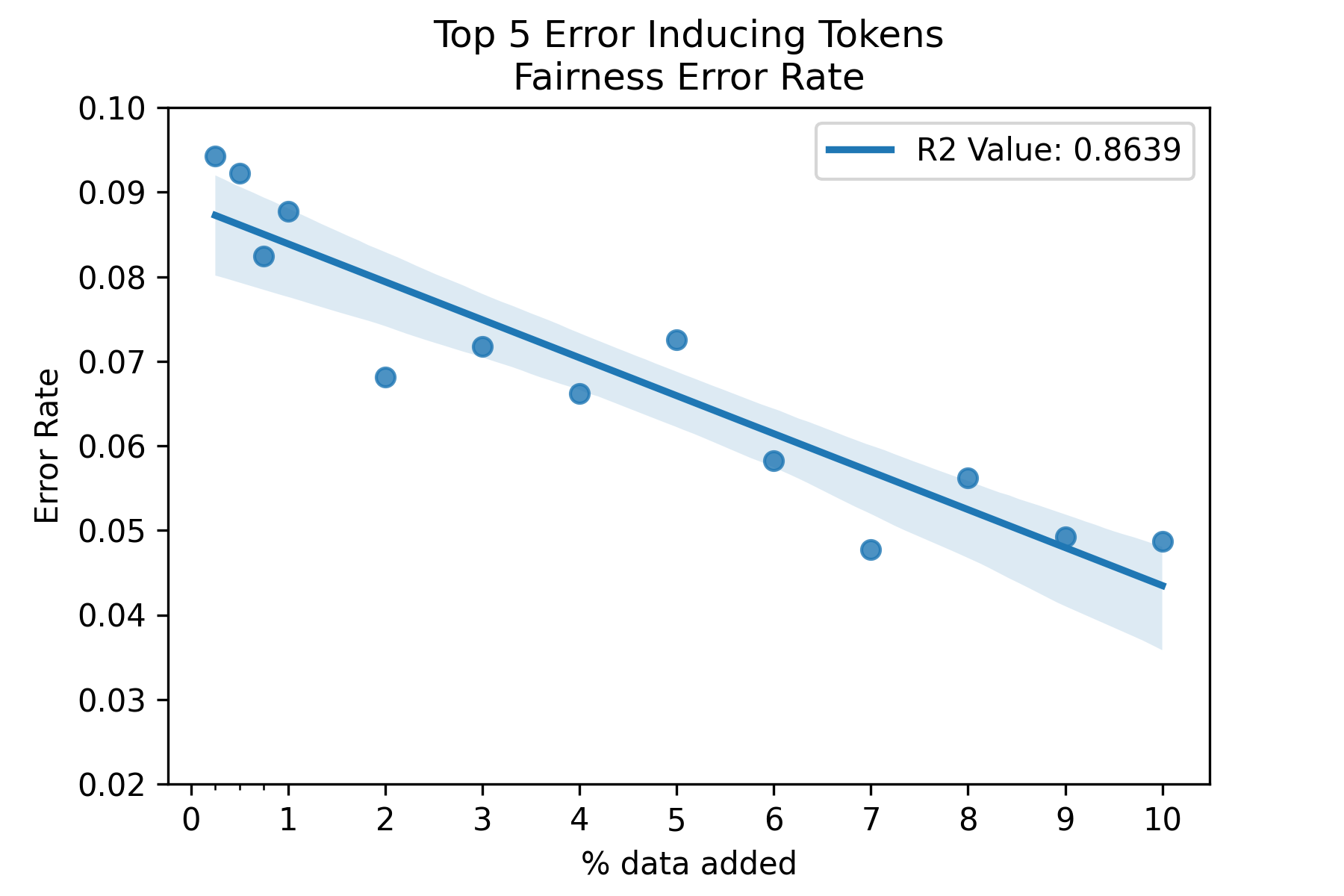}
\qquad  \qquad  \qquad 
\includegraphics[scale=0.5]{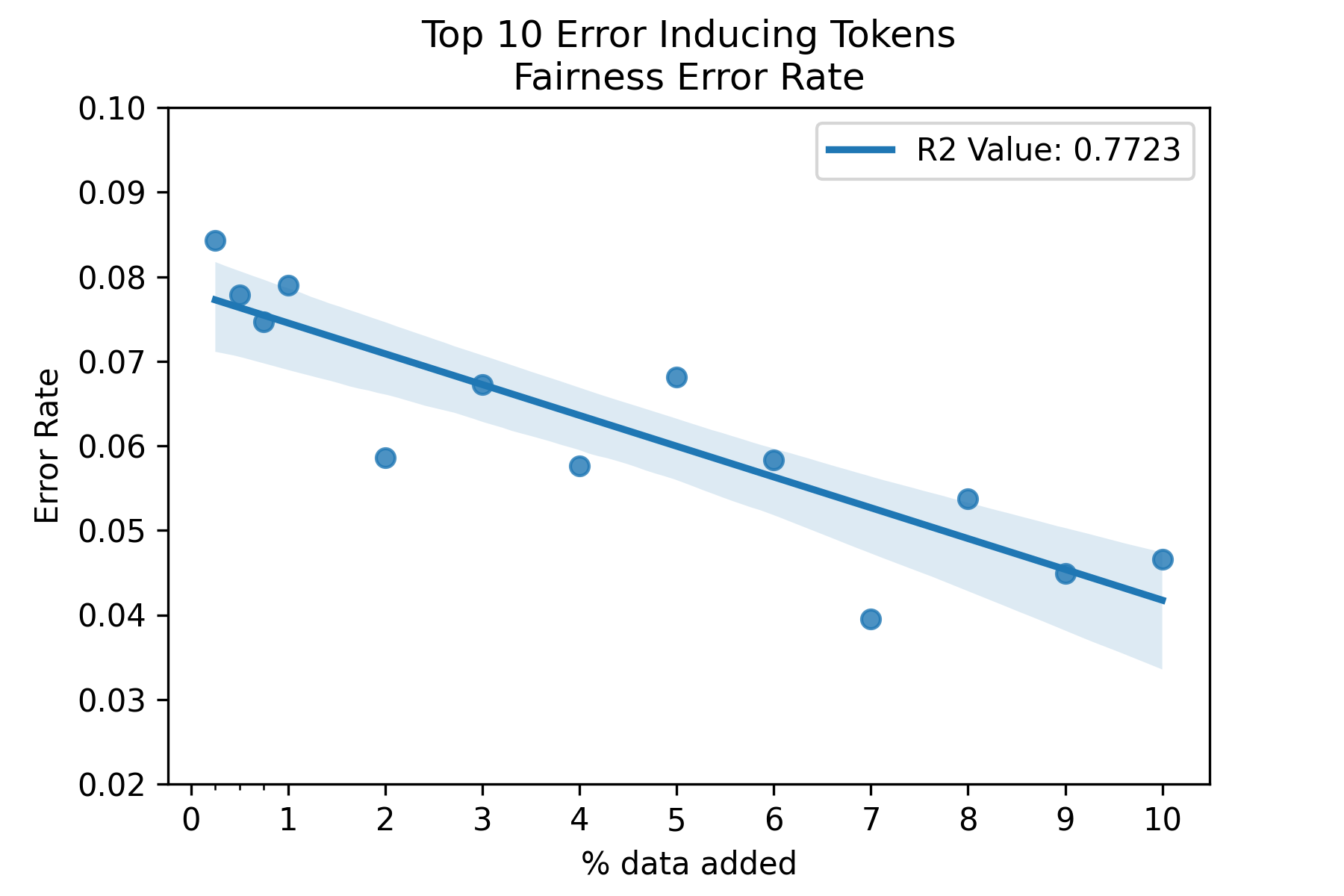}
\end{tabular}\\
\quad \quad 
(a) Fairness Error Rate for the Top 5 error tokens 
\quad \quad \quad \quad \quad \quad 
(b) Fairness Error Rate for the  Top 10 error tokens 
\quad \quad \quad \

\begin{tabular}{c}
\includegraphics[scale=0.5]{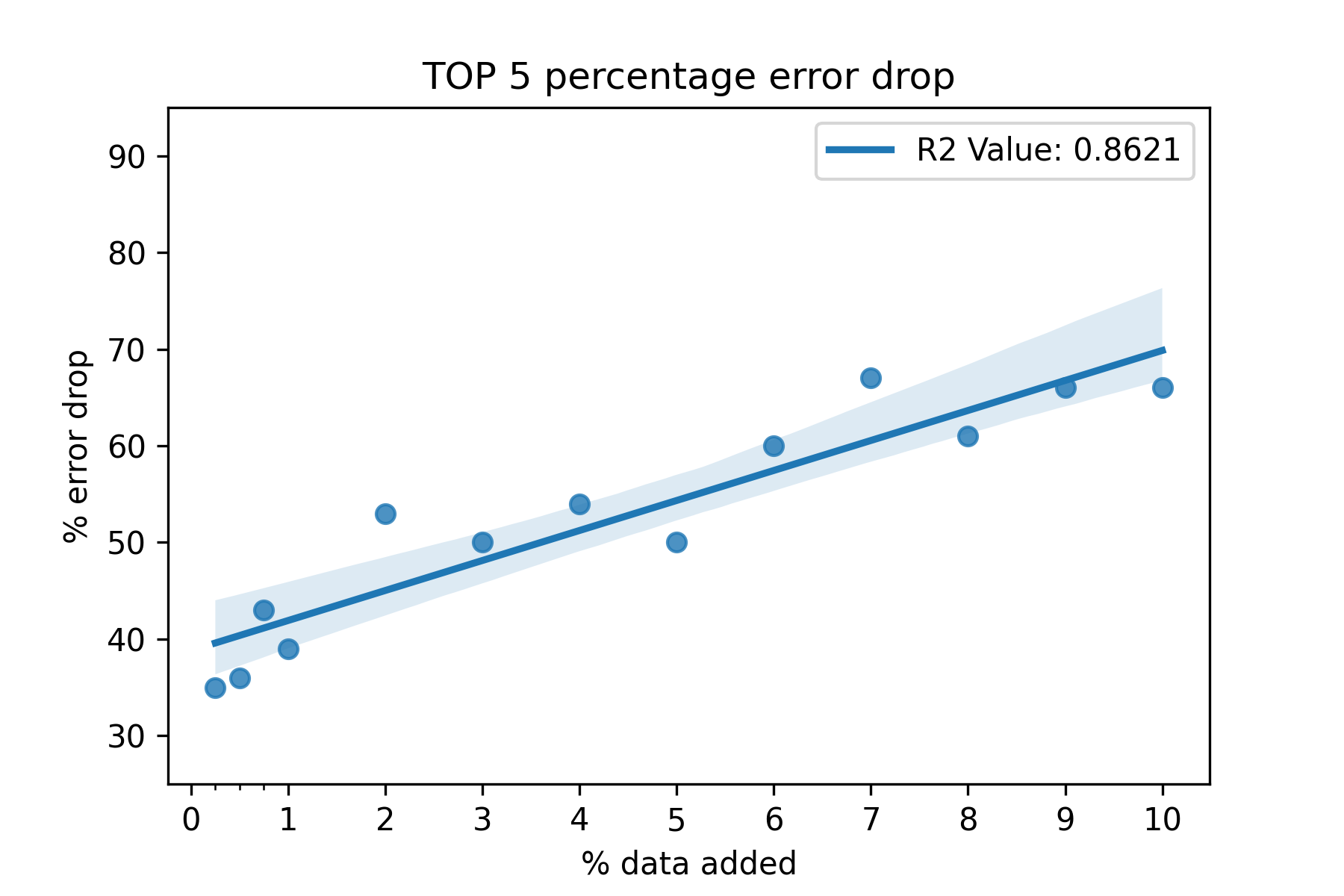}
\qquad  \qquad  \qquad 
\includegraphics[scale=0.5]{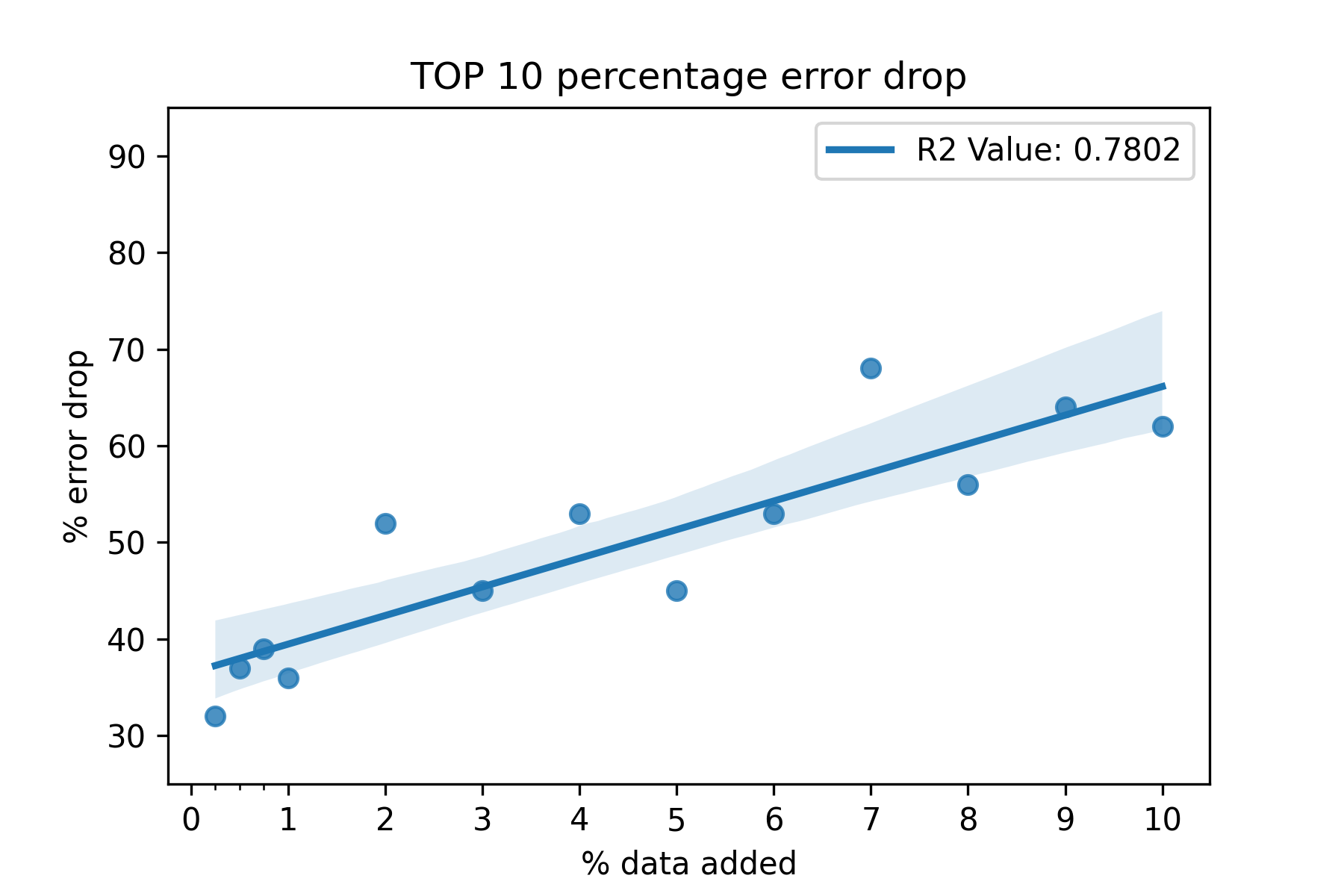}
\end{tabular}\\
(c) Drop in Fairness Violations for the Top 5 error tokens 
\quad \quad 
(d) Drop in Fairness Violations for the Top 10 error tokens 
\quad \quad \quad \
\vspace{-0.15in}
\end{center}

\caption{\small \reviseNew{
Generalisability of \NLT 's bias mitigation on unseen input sentences (from \textsc{Winogender}~\cite{winogender}) containing the top five and top 10 error-inducing input tokens, using 
re-trained models augmented with \NLT test inputs of sizes \{0.25, 0.5, 0.75, 1-10\}\% of the original training data. Charts (a) and (b) show the trend in fairness error rate of the re-trained models for each data augmentation sample size, and charts (c) and (d) show the percentage drop in fairness violations between the original model and each data augmentation sample size for re-trained models. For each chart, we compute the linear regression trend line (thick blue line) and report the R-squared value ($R^2$) (or goodness of fit). The blue shaded region surrounding the trend line is the 95\% confidence interval. 
}} 
\label{fig:gen-repair}
\end{figure*} 

\smallskip\noindent
\textbf{RQ4 Generalisability of \revise{\NLT's Bias Mitigation}:} 
\revise{
In this experiment, we examine whether \NLT 's \revise{bias mitigation (\textit{i.e.}, its 
data augmentation with 
error-inducing input tokens, and re-training with sentences containing such tokens)}
generalises to unseen input sentences, in particular, sentences in the wild that contain previously error-inducing tokens. For instance, if \NLT identified the token ``CEO" as the most error-inducing token for a sentiment analyser, we check if other sentences in the wild containing ``CEO" token still lead to fairness violations in the re-trained models obtained via \NLT's bias mitigation}. 
%
To address this, we collected five (5) and ten (10) of the topmost error-inducing input tokens identified by \NLT. 
As an example, we choose the top five or 10 most biased (fe)male occupations from our sentiment analysis experiments in RQ3. Then, using the sentences provided by a different sentiment analysis dataset \textsc{Winogender}~\cite{winogender}, we replaced these error-inducing tokens in these sentences and test them on both the original and re-trained models. \revise{
We performed model re-training using a setup similar to that of RQ3, i.e., we re-trained models with different levels of data augmentation using the Tensorflow Text Classifier Hub model. We experimented with 13 data augmentation configurations of sample sizes 0.25\% to 10\%, specifically,  \{0.25, 0.5, 0.75, 1, 2, 3, 4, 5, 6, 7, 8, 9, 10\} (\textit{see} \autoref{tab:gen-of-repair}). To mitigate the randomness in the sampling of newly generated test inputs, we sampled test inputs 10 times and trained 10 models for each configuration, following the standard for random test experiments~\cite {arcuri2011practical}. Overall, we trained 130 models, 10 models for each of the 13 data augmentation levels.}
\autoref{tab:gen-of-repair} and \autoref{fig:gen-repair} illustrate the observed 
fairness violations on our (re-)trained models, when fed with unseen inputs containing \NLT's identified error-inducing tokens. 

\revise{
In our evaluation, \emph{\NLT 's \revise{mitigation} generalises to unseen input sentences containing our diagnosed 
error-inducing input tokens}, \textit{i.e.}, unseen inputs refer to input sentences that differ from the original error-inducing inputs, 
the training data and the augmented data. \autoref{fig:gen-repair}(a) and (b) show the drop in the rate of fairness violations induced by unseen inputs in  
the re-trained models, for both the topmost five and ten error-inducing input tokens. Likewise, \autoref{fig:gen-repair}(c) and (d) show the reduction in fairness violations between the original model and the trained model for each data augmentation configuration. 
Overall, we observed that \NLT's \revise{mitigation} reduced the rate of fairness violations in a re-trained model by \revise{51}\%, on average (\textit{see} \autoref{tab:gen-of-repair}). This result suggests that \NLT 's bias mitigation generalises to unseen inputs containing previously error-inducing inputs, even when the inputs are different from the input sentences generated by the grammar, the original training data or the augmented training data. Indeed, there are error-inducing input tokens that generally induce fairness violations in sentences regardless of the task-specific input (grammar), and  \NLT can identify and mitigate against such tokens via data augmentation.
}

\begin{result}
\revise{
\NLT 's \revise{mitigation} generalises to unseen inputs: 
It reduces fairness violations in unseen inputs by about half (\revise{51}\%), on average.
} 
\end{result}

\revise{
In addition, \textit{we observed that re-training with 
the topmost five (5) error-inducing tokens outperforms re-training with the topmost 10 error-inducing tokens}. 
Specifically, for all data augmentation configurations (except 0.5\%), re-training with the  topmost five error-inducing input tokens outperformed re-training with the topmost 10 error-inducing tokens. For each data augmentation configuration, models trained with the topmost five error-inducing inputs reduced fairness violations better than models trained with the topmost 10 error tokens. Overall, re-training with the topmost five error-inducing input tokens is 5.67\% better than re-training with the topmost 10 error-inducing input tokens, on average. This is also evident by the slight difference in the trend line and $R^2$ of the top five versus top 10 error tokens (\textit{cf.} \autoref{fig:gen-repair}). Specifically, the $R^2$ value of the top five error tokens (0.8639 and 0.8621), is higher than that of the top 10 error tokens (0.7723 and 0.7802). 
This result demonstrates the efficacy of \NLT's identification of error tokens and the importance of ranking the input tokens causing fairness violations. 
}

\begin{result}
\revise{
For most configurations (12 out of 13), re-training with the topmost five (5) error-inducing tokens outperformed re-training with the topmost 10 
tokens (by 5.67\%), on average. 
}
\end{result}

\revise{
Finally, \textit{we observe a steady decrease in the number of fairness violations as the sample size of the augmented data used for model re-training increases}. For instance, for the topmost five error-inducing tokens, when 0.25\% 
of the training data is augmented for re-training, the observed reduction in error rate was \revise{35}\% (from 0.144 to 0.0943). 
Meanwhile, we observed almost twice (66\%) the reduction in fairness error rate (from 0.144 to 0.0487) with 10\% 
data augmentation (\textit{see} \autoref{tab:gen-of-repair} and \autoref{fig:gen-repair}). 
This is also evident in the trend lines, showing that there is a strong positive trend ($R^2 > 0.7$) for all charts (\textit{see} \autoref{fig:gen-repair}(c) and (d)). 
This result implies that the reduction in fairness error rate improves as the size of the augmented training data increases \textit{even} for unseen inputs, \textit{i.e.}, our results on fairness improvements (reported in RQ3) generalise to unseen inputs.
}

\begin{result}
\revise{
For unseen inputs, there is a steady reduction in fairness violations as the size of the augmented re-training data 
increases.
} 
\end{result}


\begin{table*}[]
\centering
\caption{Grammar Coverage achieved by \NLT RAND versus PROB for each/all tasks, showing the RAND coverage in normal text and PROB coverage in parenthesis ``()", as well as the percentage reduction in grammar coverage achieved by PROB, in comparison to RAND. 
}
\label{tab:coverage-prov-vs-rand}{
\begin{tabular}{|l|r|r|r|r|r|r|r|}
\hline
\multirow{2}{*}{ \textbf{Tasks} (\#MUT)} & \textbf{Input} & \multicolumn{3}{c|}{\textbf{Terminal Symbols}} & \multicolumn{3}{c|}{\textbf{Pairwise Symbols}} \\
& \textbf{Grammar} & \textbf{\#AllSymbols} & \textbf{\#Covered }& \textbf{\% Covered}  &\textbf{ \#AllPairs }& \textbf{\#Covered }&  \textbf {\% Covered} \\
\hline
\multirow{2}{*}{Coreference Resolution (3)} & Unambiguous  & 276 & 272 (216) & 98.6 (78.3) & 1449 & 1445 (649)  & 99.7 (44.8) \\
& Ambiguous  & 369 & 369 (369)  & 100.0 (100.0) & 4416 & 4158 (3434) & 94.2 (77.8) \\\hline
Mask Language Modeling (4) & Ambiguous  & 284 & 284 (276) & 100.0 (97.2) & 1848 & 1842 (1472)  & 99.7 (79.7) \\
\hline
Sentiment Analysis (11) & Ambiguous  & 2497 & 2464 (2211)  & 98.7 (88.5) & 89089 & 14366  (11979) & 16.1  (13.4) \\
\hline 
 \multicolumn{2}{|l|}{\textbf{Overall}} & 3426 & 3389 (3072) & 98.9  (89.7) & 96802 & 21811 (17534) & 22.5 (18.1) \\
\hline
 \multicolumn{4}{|l|}{\textbf{Percentage Reduction in Coverage (\%)}}  & 9.4 & \multicolumn{3}{r|}{19.6}  \\
\hline
\end{tabular}}
\end{table*}

\smallskip\noindent
\textbf{RQ5 Effectiveness of test optimisation:}
We investigate the effectiveness of our test optimisation approach, 
i.e., the probabilistic test generator (PROB). In particular, we 
examine the  effectiveness of \NLT's PROB, in comparison to the 
random test generation (RAND) (reported in RQ1). 
We also compare the \textit{grammar coverage} achieved by both RAND and PROB, in order to determine whether PROB's test optimisation achieves a higher error rate whilst covering fewer grammar production rules, in terms of terminal nodes and pairwise sensitive terminals. 

\NLT's probabilistic test generation approach (PROB) outperforms 
the random generator (RAND), in terms of the number of individual 
fairness violations found and the total number of generated test 
cases. Specifically,  PROB triggered 54\% (1370) more unique fairness 
violations in comparison to RAND, on average 
(\textit{see row ``Average" in} \Cref{tab:results-summary-ind-fairness}). 
In addition, PROB reduced the total number of generated test cases by 
10\% (\textit{see row ``Total" in} \Cref{tab:results-summary-ind-fairness}). 
Consequently, \NLT's PROB induced a higher failure rate (61\% more) than RAND, 
for individual fairness violations (\textit{see row ``Average" in} 
\Cref{tab:results-summary-ind-fairness}). These results show the improvement 
in test generation effectiveness 
of our fairness test optimizer (PROB).

\begin{result}
PROB exposed 54\% more unique individual fairness violations than RAND.
\end{result}

In our evaluation, \NLT 's probabilistic test generation approach (PROB) achieves less grammar coverage than the random generator (RAND), even though, PROB reveals more fairness errors than RAND. Overall, \NLT 's PROB covered nine percent fewer terminal nodes than RAND, and about 20\% fewer pairwise sensitive terminal nodes (\textit{see} \autoref{tab:coverage-prov-vs-rand}). This result demonstrates that PROB is able to reveal more unique errors, despite covering fewer (error-inducing) terminal nodes, suggesting that \NLT
(\textit{i.e.,} PROB) accurately learns the error-inducing input tokens necessary to induce more unique fairness errors. 

\begin{result}
\revise{
PROB achieves lower 
grammar coverage than RAND, despite revealing more (54\%) 
fairness errors.
} 
\end{result}

%
%
%
%


\begin{table}[]
\centering
\caption{Comparative Effectiveness: \NLT versus \Chk, the higher error rates and positive improvements are marked in \textbf{bold}}
\label{tab:astraea-vs-checklist}{
\begin{tabular}{|l|r|r|r|}
\hline
& \multicolumn{2}{c|}{\textbf{Error Rates}} & \textbf{\NLT \%} \\
 \textbf{MUT} & \textbf{\NLT} & \textbf{\Chk} & \textbf{Improvement} \\ 
 & (\#errors) &  (\#errors) & (\#folds) \\ 
\hline
\textbf{Stanford CoreNLP} & \textbf{0.08 (1416)} & 0.05 (656) & \textbf{51(1.5)} \\
\textbf{VaderSentiment}  & \textbf{0.11 (1574)} & 0.04 (442) & \textbf{222 (3.2) }\\
\textbf{NLTK-Vader} & \textbf{0.10 (1319)} & 0.04 (442) & \textbf{190  (2.9)} \\
\textbf{Google NLP} & \textbf{0.12 (2172)} & 0.11 (1344) & \textbf{13 (1.1)} \\
\textbf{TextBlob I} & 0.08 (1131) & \textbf{0.23 (2846)} & -67 (0.3) \\
\textbf{TextBlob II} & 0.08 (1138) & \textbf{0.23 (2846)} & -67 (0.3) \\
\hline
\end{tabular}}
\end{table}

\begin{figure}[t]
\begin{center}
\begin{tabular}{c}
\includegraphics[scale=0.5]{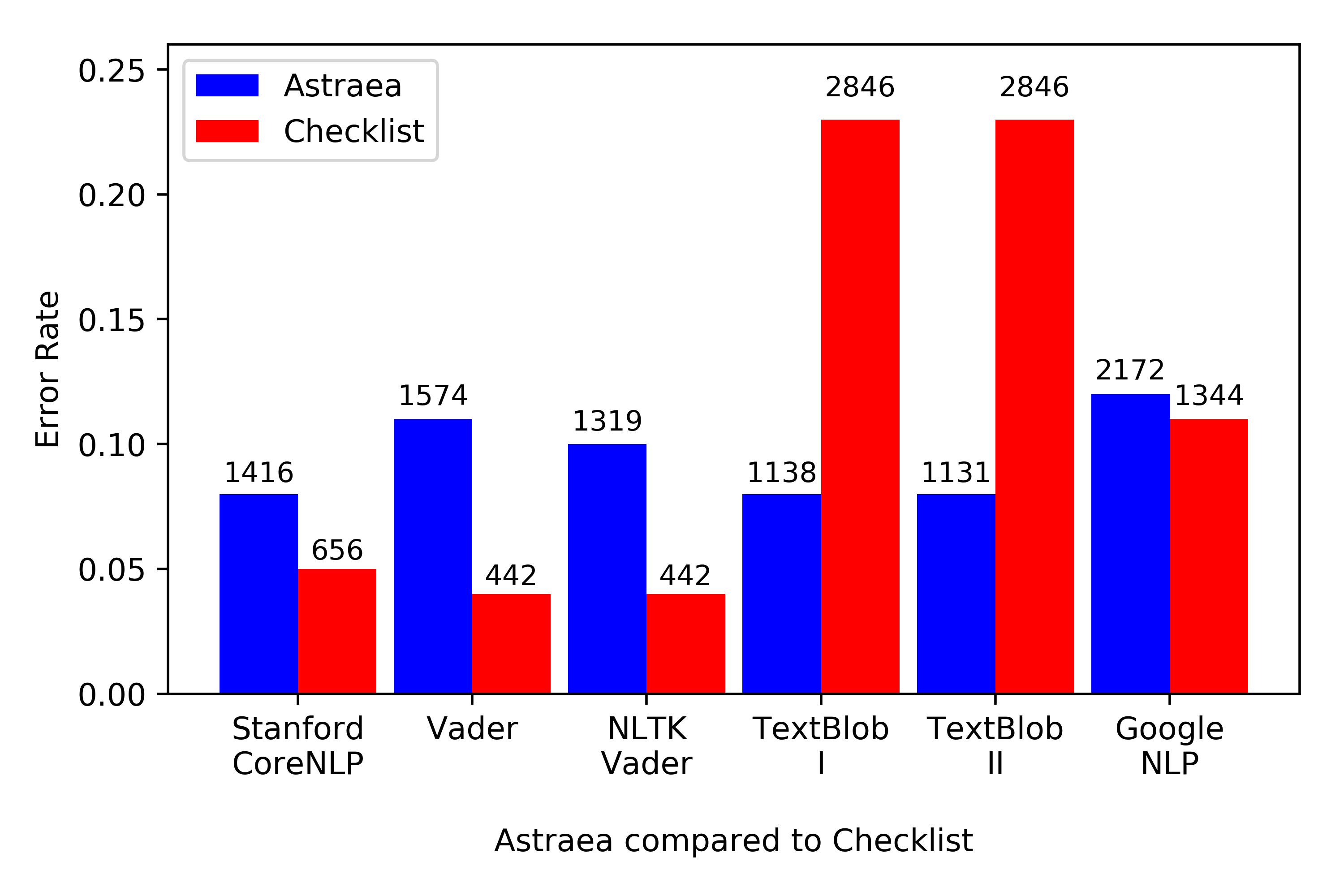}
\end{tabular}
\vspace{-0.15in}
\end{center}
\caption{\small \NLT versus \Chk: Comparing the fault revealing effectiveness of \NLT to that of \Chk w.r.t. to the error rate, numbers atop the bars indicate the number of fairness errors found}
\label{fig:astraea-vs-checklist}
\end{figure}

\smallskip\noindent
\textbf{RQ6 Comparative Effectiveness:} 
We compare the effectiveness of \NLT to the state of the art in NLP testing,\textit{ i.e.}, \textsc{Checklist} 
and \textsc{MT-NLP}. \textsc{Checklist} is a schema based NLP testing approach that generates valid inputs to improve the performance of NLP systems~\cite{checklist}, and 
\textsc{MT-NLP} is perturbation-based fairness testing approach for sentiment analyzers. In this experiment, we compare the effectiveness of \NLT to both approaches in revealing fairness violations. In particular, we compare the number of fairness violations revealed by each approach when we feed its generated sentences to 
each pre-trained sentiment analyzer in our dataset. \autoref{tab:astraea-vs-checklist} and \autoref{fig:astraea-vs-checklist} illustrate the comparative effectiveness of \NLT and \Chk on all (6) pre-trained sentiment analyzers in our setup. 

Our evaluation results show that \NLT 
had a higher error rate than 
\textsc{Checklist} for all pre-trained models, except for the two TextBlob sentiment anlayzers (\textit{i.e.}, TextBlob's Naive Bayes and Pattern analysis models).  \autoref{fig:astraea-vs-checklist} shows that for most (four of six) of our subjects, \NLT had between 13 to 222 percent higher error rate than 
\textsc{Checklist}. Specifically, \NLT had more than three times the error rate of \Chk for both Vader and NLTK-vader sentiment analyzers, and twice the error rate of \Chk for Stanford Core NLP (\textit{see} \autoref{tab:astraea-vs-checklist}). Meanwhile, \Chk outperformed \NLT for the TextBlob models where \NLT had only a third of the error rate of \Chk. 
This result suggests 
\NLT is more effective across our subjects, and is complementary to \Chk for revealing fairness errors.

\begin{result}
\NLT had a higher error rate than \Chk for most (4/6) of our subject programs. 
\end{result}

Additionally, we compare the effectiveness of \NLT to that of previous work --- \Mt~\cite{ijcai_nlp_testing}. \Mt is a perturbation-based fairness testing approach for NLP systems. We compare the performance of \NLT and \Mt on the subject program used in Ma et al.~\cite{ijcai_nlp_testing}, \textit{i.e.}, the popular Google NLP sentiment analyzer. Both \NLT and MT-NLP~\cite{ijcai_nlp_testing} were evaluated using Google's sentiment analysis engine.

Our results show that \NLT is more effective than \Mt in terms of the number of fairness errors found and error rate. \NLT 's RAND and PROB revealed 14 and 16 times as many fairness violations as \Mt, respectively. In total, \NLT found 2,172 fairness violations (out of 17,713 generated inputs)  for PROB and 1,893 violations (out of 17,711 generated inputs)  for RAND (\textit{see} Google NLP in \autoref{tab:astraea-vs-checklist}). Meanwhile, \Mt found 140 fairness errors out of 30,504 generated inputs~\cite{ijcai_nlp_testing}. \NLT also outperforms \Mt in terms of error rate, it discovers fairness errors at a rate that is 23 and 26 times higher than that of \Mt, for \NLT 's RAND and PROB, respectively.  \revise{Clearly, these results show that \NLT is more effective than the perturbation-based fairness testing of \Mt.}

\begin{result}
\NLT reveals fairness errors at a rate that is up to 26 times higher than that of \Mt.
\end{result}

\smallskip\noindent 
\textbf{RQ7 Stability of \NLT's test generation: }
To illustrate the stability of \NLT, we examine the impact of randomness 
on the effectiveness of \NLT for both \NLT (RAND) and \NLT (PROB). 
We compared results for ten runs of \NLT for the Coreference NLP tasks. 
In this evaluation, we tested for  gender bias in three MUTs, 
namely, Allen NLP, Neural Coref and Stanford CoreNLP.

Overall, our evaluation reveals that \NLT is stable in terms 
of discovering fairness violations and the number of generated test 
cases. Across all runs, \NLT had a very low standard deviation (SD). 
In terms of error rate, \NLT had an SD of 0.0054, on average. 
Specifically, in the RAND mode, \NLT had an SD of 0.0045, and in 
the PROB mode, the SD was 0.0063. This demonstrates the 
\textit{negligible effect} of randomness on \NLT's effectiveness. 
\revise{
Specifically, the inherent randomness in \NLT has little 
impact on the number of fairness violations found or the error rate.
}

\begin{result}
\NLT is stable, the effect of randomness on \NLT's effectiveness is negligible. 
\end{result}

\begin{table}[]
\begin{center}
\caption{Syntactic validity of Generated Inputs}
\label{tab:grammar_correctness}

\resizebox{\linewidth}{!}{
\begin{tabular}{|l|l|l|l|l|}
\hline
\textbf{NLP Task}                                                                      & \textbf{Bias}                                                           & \textbf{Score} & \textbf{Alerts} & \textbf{\begin{tabular}[c]{@{}l@{}}Grammar \\ Errors\end{tabular}}                                                                                            \\ \hline
\multirow{4}{*}{\textbf{Coref.}}                                                       & \textit{Gender Amb.}                                                    & 97             & 12              & laborer -\textgreater labourer                                                                                                                                \\ \cline{2-5} 
                                                                                       & \textit{Gender Unamb.}                                                  & 97             & 6               & laborer -\textgreater labourer                                                                                                                                \\ \cline{2-5} 
                                                                                       & \textit{Religion}                                                     & 97            & 2             & laborer -\textgreater labourer                                                                                                                                \\ \cline{2-5} 
                                                                                       & \textit{Occupation}                                                       & 96             & 5               & laborer -\textgreater labourer                                                                                                                                \\ \hline
\textbf{MLM}                                                                           & \textit{Occupation}                                                     & 97             & 22              & \begin{tabular}[c]{@{}l@{}}laborer -\textgreater labourer, \\ neighborhood -\textgreater\\ neighbourhood,\\ counselor -\textgreater\\ counsellor\end{tabular} \\ \hline
\multirow{6}{*}{\textbf{\begin{tabular}[c]{@{}l@{}}Sentiment\\ Analysis\end{tabular}}} & \textit{\begin{tabular}[c]{@{}l@{}}Gender\\ (Direct)\end{tabular}}      & 99             & 2               & feel -\textgreater feels                                                                                                                                      \\ \cline{2-5} 
                                                                                       & \textit{\begin{tabular}[c]{@{}l@{}}Gender\\ (Random)\end{tabular}}      & 99             & 1               & \begin{tabular}[c]{@{}l@{}}The housekeeper -\textgreater\\ Housekeeper\\ tailor made -\textgreater\\ tailor-made\end{tabular}                                 \\ \cline{2-5} 
                                                                                       & \textit{\begin{tabular}[c]{@{}l@{}}Gender \\ (Occupation)\end{tabular}} & 99             & 2               & NA                                                                                                                                                            \\ \cline{2-5} 
                                                                                       & \textit{\begin{tabular}[c]{@{}l@{}}Gender \\ (Name)\end{tabular}}       & 100            & 0               & NA                                                                                                                                                            \\ \cline{2-5} 
                                                                                       & \textit{Race}                                                           & 100            & 0               & \begin{tabular}[c]{@{}l@{}}The The Paralegal -\textgreater\\ The Paralegal\\ The The Librarian -\textgreater\\ The Librarian\end{tabular}                     \\ \cline{2-5} 
                                                                                       & \textit{Neutral}                                             & 90             & 37              & feel -\textgreater feels                                                                                                                                      \\ \hline
\multicolumn{2}{|l|}{\textbf{Average}} & 97.4           & \revise{8.1}                                                                                                                                                                \\ \cline{1-4}
\end{tabular}}

\end{center}
\vspace{-0.3cm}
\end{table}

\smallskip\noindent 
\revise{
\textbf{RQ8 Validity of \NLT's generated inputs: } This RQ evaluates the correctness of the input grammars employed by \NLT. Specifically, this is accomplished by evaluating the syntactic and semantic validity of the resulting input sentences. 
To this end, we conducted two experiments to examine the correctness of the generated inputs and investigate how \NLT's generated inputs compare to human-written sentences, in terms of sensibility. First, we fed all generated inputs to a grammar checker (\textit{i.e.}, \texttt{grammarly}) to evaluate their syntactic validity. \autoref{tab:grammar_correctness} highlights the syntactic validity results for our generated inputs. Secondly, we conducted a user study with 205 participants to evaluate the semantic validity (\textit{i.e.}, sensibility) of \NLT's generated inputs, especially in comparison to human-written input sentences (from \textsc{Winogender}~\cite{winogender}). In this experiment, we compare the human-rated sensibility of 10 human-written sentences to \NLT's generated sentences, in particular, 
using 10 benign sentences and 10 error-inducing sentences. 
\autoref{table:sem-user-study} highlights the aggregated results of our semantic validity user study. 
}

\smallskip\noindent
\revise{
\textbf{Syntactic Validity:} 
Our evaluation results show that \textit{almost all input sentences (97.4\%) generated by \NLT are syntactically valid}. \Cref{tab:grammar_correctness} highlights the correctness of \NLT 's generated inputs, it shows that the  majority (97.4\%) of the generated sentences are syntactically valid. In addition, we employ \texttt{Grammarly} to evaluate the correctness, clarity, engagement and delivery of \NLT's generated input sentences. 
Our evaluation results showed that the clarity, engagement and delivery of \NLT's generated sentences are ``{\em very clear, just right} and {\em very engaging}''. 
\texttt{Grammarly} recommended very few corrections for our generated sentences, in particular, correctness alerts were low at about \revise{8.1} alerts on average (\textit{see} \autoref{tab:grammar_correctness}). 
Common errors found in the generated sentences 
can be easily corrected by updating the terminal symbols, more importantly, these errors do not impact fairness checks. Found syntactic errors include errors about English dialects (American versus British English, \textit{e.g.} ``laborer" vs ``labourer''), minor grammar errors (``feel" vs ``feels'') and accidental incorrect terminal symbols (``The The paralegal'' vs ``The paralegal''). 
Overall, this result suggests that the inputs generated by \NLT are mostly syntactically valid, and the input grammar employed for this generation are syntactically correct. 
}
\begin{result}
\revise{
Most input sentences (97.4\%)  generated by \NLT are syntactically valid.
}
\end{result}



\smallskip\noindent
\revise{
\textbf{Semantic Validity:} We conducted a user study to evaluate the 
semantic validity (sensibility) of the
input sentences generated by 
\NLT\footnote{The user study form can be seen here: 
\url{https://forms.gle/bEudnfucckPkG8GP6}}.
In this experiment, we randomly selected 10 sentences from the \textsc{Winogender} dataset~\cite{winogender} and 20 sentences from 
the inputs 
generated by 
\NLT, specifically, ten benign sentences and ten error-inducing sentences. The handcrafted, human-written \textsc{Winogender}~\cite{winogender} sentences are chosen as a 
baseline for sensibility of input sentences, such that we can compare the sensibility \NLT's automatically generated sentences to human-crafted sentences. 
In total, 
we had 30 sentences for the user study, we provide all sentences in a random order to participants while posing the  
following question: 
}
\begin{center}
	{\fontfamily{qcr}\selectfont 
\revise{
How sensible are these sentences on a scale of one (1) to 10 (one being completely nonsensical, 10 being perfectly sensible)?
}
}
\end{center}
\revise{
For each sentence in the survey, we asked participants 
to rate its sensibility  using a 10 point Likert scale. 
Each sentence was rated from one (1) to 10, with score one being completely 
nonsensical, and 10 meaning perfectly sensible. 
The study had 205 participants recruited 
via Amazon Mechanical Turk (mTurk or AMT). 
The user study took approximately five hours and a study participants took about 7 minutes and 29 seconds to complete the survey, 
on average. 
}

\revise{Our evaluation results showed that \textit{\NLT's generated sentences
are mostly sensible (6.3/10), and comparable to human-written sentence, they were rated \revise{81}\% as sensible as the handcrafted sentences from the \textsc{Winogender} dataset~\cite{winogender}}, on average. 
The set of (20) input sentences generated by \NLT had a \revise{6.3} sensibility score, while human-written input sentences (from \textsc{Winogender}) had a \revise{7.8} sensibility score, on average. \autoref{table:sem-user-study} 
highlights the sensibility score for each set of sentences employed in our user study. 
In particular, the error-inducing sentences generated by \NLT were rated mostly sensible, even slightly more sensible than the benign sentences. Notably, \NLT's error-inducing inputs were rated \revise{83}\%  as sensible as human-written sentences (\textit{i.e.,} \textsc{Winogender}). The error-inducing sentences are also slightly more sensible (\revise{7}\%) than the benign sentences generated by \NLT, with benign sentences rated \revise{6.1} versus 
error-inducing sentences rated \revise{6.5}, on average (\textit{see} \autoref{table:sem-user-study}). This result suggests that the fairness violations induced by \NLT are from sensible and human-comprehensible sentences. 
}

\begin{result}
\revise{
\NLT 's generated sentences are mostly sensible (6.3/10) and 
almost (\revise{81}\%) as sensible as 
human written sentences.
}
\end{result}

%
%
%


\begin{table}
\caption{Semantic User Study Scores}
\centering
{\scriptsize
\begin{tabular}{lrrrr}\toprule
& \makecell{Winogender\\(Baseline)} &	\makecell{Astraea\\(Overall)} &	\makecell{Astraea\\(Error)} &	\makecell{Astraea\\(Non-Error)} \\ \cmidrule(lr){1-5}
\multicolumn{1}{l}{Mean} &	7.848&	6.323&	6.529& 	6.118\\
\textit{\% drop} & - & 19.42\%	& 16.81\% &	22.04\% \\ \midrule

\multicolumn{1}{l}{Median} & 	8.50 &	6.75 &	7.00 & 	6.50\\
\textit{\% drop} & - & 20.59\%	& 17.65\% &	23.53\% \\ \midrule
\label{table:sem-user-study}
\end{tabular}
}
\end{table}

\subsection{Discussions and Future Outlook}
\label{sec:discussion-future}









\smallskip\noindent
\revise{
\textbf{Ethical Considerations:} In this section, we consider the ethical issues related to the use of \NLT. In particular, the ethical implications of applying \NLT in analyzing and mitigating societal biases, as well as issues relating to the application of \NLT  in fairness testing, especially on marginalized individuals and (minority) groups.
}

\smallskip\noindent
\revise{
\textit{Intended Use: } 
The \textit{intended use} of \NLT is to analyse, detect and mitigate undesirable biases in text-based NLP tasks. Although, test generation is important to ensure software fairness for ML-based software systems, it is pertinent to note that it is not sufficient to address the problem of fairness in (NLP-based) software systems. Our recommendation is that tools such as \NLT should be deployed as part of the ML pipeline and end-to-end analysis to validate fairness properties. 
Indeed, \NLT should be deployed within the context of a well-defined societal or institutional fairness policy. 
Besides, there are other concerns when applying software systems (such as \NLT) to ensure fair and inclusive ML systems. Notably, it is important to define the social context of \NLT's application, the ethical concerns in terms of the societal biases in consideration, the desirable bias policy and the intended use cases of the NLP system at hand. These concerns inform fair and inclusive design and analysis of NLP systems. 
For more details on the ethical concerns for NLP systems, Hovy and Spruit ~\cite{hovy2016social} provide a comprehensive survey on the social impact of NLP systems, especially in terms of their impact on social justice,
\textit{i.e.}, equal opportunities for individuals and groups.
}

\smallskip\noindent
\revise{
\textit{Experimental Design: }
In this paper, we have evaluated \NLT on a range of NLP tasks where it demonstrates considerable benefit in improving software fairness. Beyond fairness testing, we believe our implementation and evaluation of \NLT has very limited harmful applications. \NLT's application does not risk deterring fairness, amplifying bias, or enabling ethical or security issues such as privacy and unintended release of personal information or identities. 
In our experiments, we ensure to avoid attributes that involve \textit{normative judgments} that may cause harm when interpreted pejoratively~\cite{denton2019detecting}. For instance, our design of \NLT ensures that generated input sentences 
are strictly 
tied to the allowed values in the production rules for a specific noun or pronoun, without prejudice for specific attributes. 
In addition, in our experiments, we have employed data sets, input grammars and programs that have been made publicly available 
to allow for scrutiny, reuse and reproducibilty.
}

\smallskip\noindent
\revise{
\textit{Non-binary gender:} In our experiments, we have only considered binary genders due to the limitations of our subject programs. Indeed, most NLP tools do not account for non-binary genders ~\cite{sun2019mitigating, hovy2016social, zhao2017men}. 
However, we do not condone the classification of gender into binary categories. We believe such rigid classifications may be harmful, especially to minority groups and marginalized individuals. 
The aim of this work is not to perpetuate such stance or reinforce rigid social classifications of gender. In fact, \NLT allows to test for non-binary gender, \textit{e.g.},  by adding their corresponding non-binary nouns or pronouns to the input grammar. This is evident in our preliminary experiments testing non-binary genders on our subject programs. For instance, we found that all of the coreference resolution (coref) systems in our setup do not recognize or account for non-binary gender (\textit{e.g.}, the singular {\em they}). Specifically, over 90\% of the sentences generated by \NLT (containing a singular {\em they} as a sensitive attribute) for all coreference resolution  (Coref) subject programs do not yield any output. Thus, in our experiments, we did not evaluate for non-binary gender, but we note that as NLP systems improve to be more gender inclusive, \NLT can be trivially extended to test for non-binary gender. Overall, similar to the stance of the research community~\cite{denton2019detecting}, we 
believe that gender attributes should not be binary or considered perceptually obvious and discernible in text-based systems.
}
%
%


\smallskip\noindent
\revise{
\textit{Age-related Bias:} In our evaluation, we have not considered other poorly understood biases (\textit{e.g.}, age-related biases). However, we expect \NLT to perform well on such poorly understood biases provided there are input tokens that characterize the societal bias. 
Extending an existing input grammar for such biases is trivial provided there is a well-defined age bias policy and a corresponding set of age-related input tokens to be added to the input grammar. As an example, fairness analysis for age-related bias for our NLP tasks can be characterized by adding age-related adjective as terminals, \textit{e.g.}, ``young'' versus ``old'',  or ``teen'' versus ``aged''. This is similar to how D{\'\i}az et al.~\cite{diaz2018addressingAgeBias} addressed age-related biases. To demonstrate this, we adapt one of our grammars used for co-referencing to 
encode age-related biases for the co-reference analysis task.\footnote{See here: \url{https://bit.ly/3s0PKKF}}
It is important to note that 
this is just a {\em demonstration} 
and the actual efficacy of \NLT's performance on 
such biases is not comprehensively understood. In the future, we plan to study such poorly understood societal biases, we also encourage other researchers to investigate approaches to analyse and mitigate 
such under-studied societal biases. 
}


\smallskip\noindent
\revise{
\textbf{Test Generation:} Let us discuss the issues related to the testing methodology of \NLT, in particular, the \textit{choice of test oracle} and the potential to generate \textit{redundant test cases}. 
}

\smallskip\noindent
\revise{
\textit{Alternative Test Oracles:} 
In this paper, we have employed metamorphic test oracles to compare the outputs of similar discriminatory input sentences. \NLT requires this oracle to automatically detect violations. For improving fairness, we require a dataset to augment our training data set. This augmented dataset is then used for re-training. To achieve this, we employ a predictive oracle to determine the ground truth output label for the newly generated test inputs used in our data augmentation. Our predictive oracle is based on the input grammar, in particular it checks for the presence of certain terminals in the generated inputs (as described in \autoref{sec:results}.1). It is important to note that this oracle is simple and it is only sound with respect to the input grammar, indeed it is not sound in general for any input sentence or grammar. Indeed, defining our rule-based oracle for a large, complex or highly expressive input grammar may be very difficult, incomplete and impact the expressiveness of the input grammar and the resulting input sentences. This is in particular a very difficult problem~\cite{barr2014oracle}, especially in the absence of ground truth about the output labels of generated input sentences. Hence, it may be necessary to employ more powerful or \textit{alternative} test oracles for more expressive input sentences or grammars.
}

\revise{
Besides, there are 
alternative approaches to generate predictive test oracles. 
For instance, 
other researchers have employed probabilistic and majority voting oracles for the same purpose. In particular, \textsc{TransRepair}~\cite{sun2020automatic} employs a probabilistic majority voting oracle for inconsistency testing, by feeding several similar discriminatory inputs to a model and using the most common outcome as the ground truth. Similarly, one can employ an ensemble of models, \textit{i.e.}, by feeding a single input or several similar discriminatory inputs to these models, and taking the most common outcome as the ground truth~\cite{barr2014oracle}.
}

\smallskip\noindent
\revise{
\textit{Redundant Test cases:} 
\NLT generates input sentences by exploring the input grammar, especially in the random (RAND) exploration mode. Hence, it may generate test cases that are redundant, \textit{i.e.}, non-unique discriminatory inputs. 
For instance, \NLT may repeatedly generate a set of discriminatory input sentences exposing similar fairness violations. 
To mitigate this, \NLT also has a more targeted phase, the PROB mode (reported in RQ5). In this phase, 
\NLT automatically generates input sentences that 
target seen fairness violations to expose more closely-related violations. This is evident in RQ5 where \NLT reduces redundant test cases. In particular, \NLT in PROB mode generates fewer unique input sentences, and yet exposed more unique fairness violations than the random exploration mode of \NLT (RAND). 
}

\revise{
In the future, we plan to investigate more targeted approaches that can reduce the number of redundant test cases, besides our probabilistic mode (PROB in RQ5). 
We plan to reduce the number of redundant tests by exploring alternative approaches, such as coverage-driven approaches (\textit{e.g.}, \textsc{OGMA}~\cite{ogma}), mutation-driven approaches (\textit{e.g.}, \textsc{TransRepair}~\cite{sun2020automatic}), 
and directed test generation approaches (\textit{e.g.}, \textsc{AEQUITAS}~\cite{aequitas}). 
}

\smallskip\noindent
\revise{
\textbf{Data Augmentation:} 
The aim of our experiments concerning model-retraining via data augmentation (in RQ3 and RQ4) is to demonstrate that our 
approach is effective in 
improving software fairness. 
Our goal is not to determine the optimal ratio of data augmentation that achieves the best mitigation of fairness violations. Even though our experiments demonstrate that as the percentage of augmented data increases, the rate of fairness violations decreases (\textit{see} RQ3 and RQ4), we do not ascertain the best data augmentation ratio for the optimal reduction in fairness violations. 
Determining the best ratio for data augmentation is a different optimization problem. In fact, 
this optimization problem requires further investigation to determine when data augmentation is sufficient to ensure maximal reduction in fairness violations. 
}

\smallskip\noindent
\revise{
\textbf{Disclaimer:} The goal of this work is not to determine the correct, desired or expected outcomes for a task, bias or program, neither is it to define the societal policy that determines the absence or presence of a violation. 
The \textit{focus of our work is to allow developers the flexibility to analyze, test and mitigate against different biases based on their use case}. In particular, based on their own defined societal or company policy, their bias of concern and their expected behavior for the task or program. We do not intend to define the societal policy for bias (\textit{e.g.}, \textit{equality} versus \textit{equity}), however, our methodology allows to check for such defined policy via the test oracle. For instance, an \textit{equality} policy can be easily checked by ensuring the outcomes for a pair of sentences are equal (\textit{e.g.}, in RQ1), and a \textit{threshold-based} policy can be easily checked by ensuring a certain threshold in the difference in outcomes is maintained (\textit{e.g.}, in RQ2). This is important because the fairness concerns of an organization may differ depending on the task, bias or policy; our focus is to allow the flexibility to test for different use cases.
}

%
%

\revise{
As an example, in the individual fairness experiments in RQ1, 
we assume that all predicted pronouns should be equally likely for all occupations while testing for occupational/gender bias. 
Meanwhile, in the case of the MLM example in RQ2, we 
allow developers to test for group fairness violations based on different threshold configurations. 
We employed a \textit{threshold-based} policy to ascertain if there is a large disparity (exceeding the defined threshold) between the \textit{his} and \textit{her} [MASK] outcomes. Likewise, we can test for \textit{equality} policy for the same MLM group fairness task, such that 
we directly compare if both outcomes are equal. In the real world, this approach translates to checking that ``all predicted outcomes should be equally likely for all occupations''.
}

\revise{
In summary, we do not intend to define the expected outcome, policy or verdict for a subject program or task. We are focused on providing 
a flexible 
approach (\textit{i.e.,} \NLT) that is easily amenable to test bias for different 
use cases, by allowing to test for several policies, fairness criteria and biases. Although, \NLT supports testing for different policies, we believe the definition of the desirable outcome or tested bias policy (\textit{e.g.}, \textit{equality} versus \textit{equity}) 
is orthogonal to our research and dependent on the use case. Indeed, automatically determining the desirable outcome based on the use case is an open problem and it should be further studied by the research community.
}

%

\section{Limitations and Threats to Validity}
\label{sec:threats-to-validity}

\smallskip \noindent
\textbf{Grammar Construction and Correctness:}
The construction of input grammars is relatively easy and the 
initial grammar was constructed by a graduate student in 30-45 mins. We 
demonstrate the ease of grammar construction by implementing a wide
range of grammars across three tasks, namely Coreference Resolution, 
Sentiment Analysis and Masked Language Modelling and for over fifteen 
models under test. Additionally, we release the Python implementations 
of the grammars for future expansion.

We attempt to construct the input grammars in such a way that the inputs 
generated by them are semantically valid (by design of the grammar). This 
is aided by the availability of 
EEC schema~\cite{kiritchenko2018examining} and 
Winogender~\cite{winogender}. To mitigate against errors that may creep 
into the grammar, we use a popular online grammar checking tool 
Grammarly~\cite{grammarly} and verify generated input correctness. 
On average, we find that the overall score is high at 97.4 (see 
\Cref{tab:grammar_correctness}). 



\smallskip \noindent
\textbf{Complex Inputs:}
\NLT's input grammars allow to specify and explore the input space 
beyond the training set. 
In our evaluation setup, it was easy to construct input grammars that expose fairness violations, within about 30 minutes. 
Our evaluation on NLP tasks with varying complexities shows that \NLT can be easily 
applied to NLP tasks.
Although grammars exist for some complex tasks (such as images\footnote{Binary format grammars for PNG and JPEG are available here: \url{https://www.sweetscape.com/010editor/repository/templates/}}), the correlation of grammar tokens and image sensitive attributes are not yet explored. This line of applications requires further research.





\smallskip \noindent
\textbf{Completeness:} %
\reviseNew{
By design, \NLT is incomplete in discovering fairness violations
due to several reasons, namely (1) \textit{input grammars} -- it can only expose the biases captured in the employed input grammar, (2) \textit{lack of guarantees in testing} -- 
in comparison to the guarantees afforded by verification, \NLT does not provide a guarantee or proof of fulfilling fairness properties and (3) \textit{finite number of generated tests} - limited number of generated inputs within a reasonable time budget.} 
\revise{Firstly, \NLT can not expose a fairness violation if the input tokens associated with the violation are not captured by the input grammar, hence, its 
effectiveness is limited by the \textit{expressiveness of the employed input grammar}. Secondly, unlike, fairness verification approaches (such as Albarghouthi et al.~\cite{albarghouthi2017fairsquare}), \NLT is a 
validation approach. Similar to typical testing approaches, it does not provide any guarantees or proof that all fairness violations have been exposed. Instead, it allows to explore the input space to assess the fairness properties of NLP systems. 
Finally, \NLT is executed for a limited number of runs, till it is saturated (\textit{i.e.}, no more unique inputs generated) or all grammar production rules are explored. 
In the former case there may be other potential fairness violations left unexposed. 
This is because the input grammar may not have been completely explored. 
}

For instance, 
\NLT runs till saturation or up to a certain number of iterations is reached. This is 
due to the absence of new unique test inputs being generated in two successive iterations. 
However, it is possible to discover more fairness violations with 
more iterations. 
For instance, this can be accomplished by extending \NLT to be grammar coverage 
driven, \textit{e.g.} via greedy exploration of all pairwise combination of (sensitive) 
terminal symbols.


\smallskip \noindent
\textbf{Generalizable ML:} 
We assume all models are generalizable to the task at hand, i.e. they should not over-fit 
to 
a specific use case or training dataset. 
\revise{
This assumption is reasonable because the input space for a model, task or use case is typically unconstrained and it is not fully captured by the training data set. 
Besides, testing for out of distribution (OOD)  inputs is necessary to ensure model reliability, i.e., validating that an ML system generalizes beyond the biases of its training dataset. 
Researchers have found that ML models can be easily fooled by out of distribution of inputs~\cite{nguyen2015deep, dola2021distribution}. 
Out of distribution (OOD) testing validates that model outputs are reliable regardless of the constraints on the training data set. Notably, Berend et al.~\cite{berend2020cats} 
investigated the importance of OOD testing for ML validation and call for the attention of data-distribution awareness during designing, testing and analysis of ML software. In their empirical study they found that distribution-aware testing is effective in improving the reliability and robustness of ML models. 
In this work, we have performed fairness testing by generating several test cases that are mostly out of distribution (\textit{i.e.}, independent of the training dataset), but necessary to ensure the reliability of the ML software. 
}

\revise{In line with the findings of Berend et al.~\cite{berend2020cats}, we design our experiments assuming that subject models generalize beyond the training dataset.} 
As an example, we expect that a sentiment analysis model trained on movie reviews, 
should generalize to other texts (e.g. conversational sentences) that express positive, 
negative or neutral emotions. 
To dampen this effect, 
we employ several models trained on varying training datasets.

\smallskip \noindent
\textbf{General Tasks:}
This refers to the generalisability of \NLT to other (NLP) tasks. 
To mitigate this threat, we evaluated 
\NLT on three distinct NLP tasks with varying complexities, 
using 18  different subjects. 
\NLT's effectiveness on all tested tasks and models shows 
it can be easily employed for other (NLP) tasks or models.

\section{Conclusion}
\label{sec:conclusion}
In this paper, we have proposed \NLT, the first grammar-based framework 
to automatically discover and diagnose fairness violations in NLP 
software. \NLT embodies a directed test generation strategy that leveraged 
the diagnosis result and it significantly improves the test effectiveness. 
Moreover, the diagnosis employed by \NLT is further used to retrain NLP 
models and significantly reduce the number of fairness errors. \NLT is designed to 
be a general fairness testing framework via an extensible grammar. This 
is validated by instantiating \NLT across three different NLP tasks comprising 
18 different models. We show that \NLT finds hundreds of thousands of 
fairness errors in these models and significantly improves software fairness 
via model re-training. \NLT provides a pathway to advance research in automated 
fairness testing of NLP software -- a crucial, yet underrepresented area 
that requires significant attention. To reproduce and further research 
activities, our tool and all experimental data are publicly available here: 
\begin{center}
\textbf{\url{https://github.com/sakshiudeshi/Astraea}}
\end{center}


\section*{Acknowledgment}
\label{sec:ack}
\reviseNew{
We thank the reviewers for their helpful comments. 
This work was partially supported by the University of Luxembourg, 
Ezekiel Soremekun acknowledges the financial support of the Institute for Advanced Studies of the University of Luxembourg through an Audacity Grant (AUDACITY-2019-Laiwyers). 
This work is also partially supported by OneConnect Financial grant number RGOCFT2001, Singapore Ministry of Education (MOE), President's Graduate Fellowship and MOE grant number MOE2018-T2-1-098.
}

\balance
\bibliographystyle{plainurl}
\bibliography{NLP-Fairness}

\end{document}